\pdfoutput=1
\documentclass[11pt,a4paper,twoside,openright,closeany,textany]{book}
\usepackage{makeidx}
\usepackage{graphicx}
\usepackage{paralist}
\usepackage{caption} 
\usepackage{amsmath}
\usepackage{amssymb}
\usepackage{mathtools}
\usepackage{fancyhdr}
\usepackage{multicol}
\usepackage{multirow}

\usepackage{mdframed}
\usepackage[usenames,dvipsnames,svgnames,table]{xcolor}
\usepackage{array}
\usepackage[%
  colorlinks=true,%
  linkcolor=black,%
  urlcolor=black,%
  citecolor=black%
]{hyperref}
\usepackage{nameref} 
\usepackage{ifthen} 
\usepackage[font={small,sl}]{caption}
\usepackage[authoryear]{natbib}
\renewcommand\bibname{References}
\newcommand{\mychapbib}{
  \addcontentsline{toc}{section}{\bibname}
  \bibliographystyle{natbib}
  \bibliography{strucbioinf}
}
\usepackage[colorinlistoftodos]{todonotes}
\usepackage{letltxmacro}

\usepackage{tocstyle} 
\usetocstyle{standard}
\settocfeature{raggedhook}{\raggedright}
\newcommand{\bibfont}{\small\raggedright}
\def\cite{\citep}
\newcommand{\citeeg}[1]{\cite[e.g.,][]{#1}}

\LetLtxMacro{\oldTodo}{\todo}
\renewcommand{\todo}[2][]{\oldTodo[#1]{TODO: #2}}

\usepackage[normalem]{ulem}

\newcommand\inwish[1]{\oldTodo[inline,color=SkyBlue]{WISH: #1}}

\newboolean{onechapter}

\newcommand{\AF}[1][~]{K.\@#1Anton#1Feenstra}
\newcommand{\SA}[1][~]{Sanne#1Abeln}

\newcommand{\HM}[1][~]{Halima#1Mouhib}

\newcommand{\QH}[1][~]{Qingzhen#1Hou}
\newcommand{\JvG}[1][~]{Juami#1H.\@#1M.\@#1van#1Gils}

\newcommand{\AM}[1][~]{Ali#1May}

\newcommand{\JG}[1][~]{\mbox{Jose}#1\mbox{Gavald\'a-Garc\'ia}}

\newcommand{\JV}[1][~]{\mbox{Jocelyne}#1\mbox{Vreede}}

\newcommand{\AR}[1][~]{\mbox{Arri\"en}#1\mbox{Symon}#1\mbox{Rauh}}

\newcommand{\orcid}[1]{\href{https://orcid.org/#1}{\raisebox{-0.7ex}{\protect\includegraphics[height=3ex]{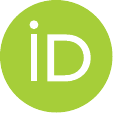}}}}
\definecolor{idgreen}{RGB}{166 206 57}
\newcommand{\mailid}[1]{\href{mailto:#1}{\raisebox{-0.3ex}{\color{idgreen}\textsf{\textbf{\Large \protect@}}}}}

\newcommand{\AFid}{\orcid{0000-0001-6755-9667}}
\newcommand{\SAid}{\orcid{0000-0002-2779-7174}}
\newcommand{\HMid}{\orcid{0000-0001-5031-3468}}
\newcommand{\JGid}{\orcid{0000-0001-6431-3442}}
\newcommand{\JvGid}{\orcid{0000-0003-3706-7818}}

\newcommand{\JVid}{\orcid{0000-0002-6977-6603}}

\newcommand{\ARid}{\orcid{0000-0001-9707-3836}}

\newcommand{\QHid}{\orcid{0000-0002-9832-4518}}

\newcommand{\AMid}{\orcid{0000-0002-0551-9966}}

\newcommand{\ACtxt}{Wrote the text}
\newcommand{\ACfig}{Created figures}
\newcommand{\ACref}{Review of current literature}
\newcommand{\ACeds}{Editorial responsibility}
\newcommand{\ACproof}{Critical proofreading}
\newcommand{\ACfb}{Non-expert feedback}

\newcommand{\Angs}[1][~]{\text{\normalfont\AA}}

\renewcommand{\and}{\quad}

\newcommand{\figlab}[1]{\textsf{\textbf{\Large #1}}}
\newcommand{\pdbref}[1]{\href{http://www.rcsb.org/pdb/explore.do?structureId=#1}{PDB:#1}}
\newcommand{\arxiv}[2][UNDEFINED]{\href{https://arxiv.org/abs/#2}{\ifthenelse{\equal{#1}{UNDEFINED}}{arxiv.org/abs/#2}{#1}}}

\newcommand{\figref}[2][]{\hyperref[fig:#2]{Figure\@~\ref*{fig:#2}#1}}
\newcommand{\tabref}[1]{\hyperref[tab:#1]{Table \ref*{tab:#1}}}
\renewcommand{\eqref}[2][]{\hyperref[eq:#2]{Equation#1\@~\ref*{eq:#2}}}
\newcommand{\panelref}[2][]{%
    \ifthenelse{\boolean{onechapter}}{%
        \hyperref[panel:#2]{Panel\@~``\nameref{panel:#2}#1''}%
    }{%
        \hyperref[panel:#2]{Panel\@~\ref*{panel:#2}#1}%
    }%
}
\newcommand{\secref}[2][n]{%
    \hyperref[sec:#2]{%
        \ifthenelse{\equal{#1}{n} }{Section\@~\ref*{sec:#2}}{}% just number
        \ifthenelse{\equal{#1}{nn}}{Section\@~\ref*{sec:#2} ``\nameref{sec:#2}''}{}% nm & nr
        \ifthenelse{\equal{#1}{N} }{``\nameref{sec:#2}''}{}% just quoted name
        \ifthenelse{\equal{#1}{NN} }{\nameref{sec:#2}}{}% just name
    }%
}
\newcommand{\chref}[2][n]{%
    \ifthenelse{\boolean{onechapter}}{%
        \ifthenelse{\equal{#2}{ChPref}     }{\arxiv[Chapter ``\nameref*{ch:#2}'']{1801.09442}}{}%
        \ifthenelse{\equal{#2}{ChIntroPS}  }{\arxiv[Chapter ``\nameref*{ch:#2}'']{1801.09442}}{}%
        \ifthenelse{\equal{#2}{ChDetVal}   }{\arxiv[Chapter ``\nameref*{ch:#2}'']{2108.02706}}{}%
        \ifthenelse{\equal{#2}{ChStrucAli} }{\arxiv[Chapter ``\nameref*{ch:#2}'']{1801.09442}}{}%
        \ifthenelse{\equal{#2}{ChDBClass}  }{\arxiv[Chapter ``\nameref*{ch:#2}'']{1801.09442}}{}%
        \ifthenelse{\equal{#2}{ChFunc}     }{\arxiv[Chapter ``\nameref*{ch:#2}'']{1801.09442}}{}%
        \ifthenelse{\equal{#2}{ChIntroPred}}{\arxiv[Chapter ``\nameref*{ch:#2}'']{1712.00407}}{}%
        \ifthenelse{\equal{#2}{ChHomMod}   }{\arxiv[Chapter ``\nameref*{ch:#2}'']{1712.00425}}{}%
        \ifthenelse{\equal{#2}{ChSSPred}   }{\arxiv[Chapter ``\nameref*{ch:#2}'']{1801.09442}}{}%
        \ifthenelse{\equal{#2}{ChFuncPred} }{\arxiv[Chapter ``\nameref*{ch:#2}'']{1801.09442}}{}%
        \ifthenelse{\equal{#2}{ChIntroDyn} }{\arxiv[Chapter ``\nameref*{ch:#2}'']{1801.09442}}{}%
        \ifthenelse{\equal{#2}{ChThermo}   }{\arxiv[Chapter ``\nameref*{ch:#2}'']{1801.09442}}{}%
        \ifthenelse{\equal{#2}{ChMD}       }{\arxiv[Chapter ``\nameref*{ch:#2}'']{1801.09442}}{}%
        \ifthenelse{\equal{#2}{ChMC}       }{\arxiv[Chapter ``\nameref*{ch:#2}'']{1801.09442}}{}%
    }{
    \hyperref[ch:#2]{%
        \ifthenelse{\equal{#1}{n} }{Chapter \ref*{ch:#2}}{}% just number
        \ifthenelse{\equal{#1}{nn}}{Chapter \ref*{ch:#2} ``\nameref{ch:#2}''}{}% name & number
        \ifthenelse{\equal{#1}{N} }{``\nameref{ch:#2}''}{}% just name
      }%
  }%
}
\newcommand{\chrefname}[1]{\hyperref[ch:#1]{Chapter \ref*{ch:#1} ``\nameref{ch:#1}''}}
\newcommand{\partref}[1]{\hyperref[#1]{Part \ref*{#1}}}
\newcommand{\appref}[1]{\hyperref[app:#1]{Appendix \ref*{app:#1}}}

\newcommand{\figsource}[1]{\protect\footnote{Figure source location: \url{#1}}}

\renewcommand{\arraystretch}{1.3}

\makeatletter \@beginparpenalty=5000 \makeatother

\newenvironment{panel}[1][]{
  \begin{figure}[htb]
    \begin{mdframed}[%
        outerlinewidth=0,%
        linecolor=CornflowerBlue!30,%
        backgroundcolor=CornflowerBlue!30,%
        innerleftmargin=14,%
        innerrightmargin=14,%
      ]
      \ifthenelse{\equal{#1}{}}{}{
        \stepcounter{panel}
		\subsection*{#1} 
      }
}{%
    \end{mdframed}
  \end{figure}
}

\newenvironment{bgreading}[1][]{
  \begin{mdframed}[%
      outerlinewidth=0,%
      linecolor=CornflowerBlue!30,%
      backgroundcolor=CornflowerBlue!30,%
      innerleftmargin=14,%
      innerrightmargin=14,%
    ]
	\ifthenelse{\equal{#1}{}}{}{
        \stepcounter{panel}
    	\subsection*{#1} 
    }
}{%
  \end{mdframed}
}

\usepackage{listings}
\definecolor{backcolour}{rgb}{0.95,0.95,0.92}
\definecolor{codegreen}{rgb}{0,0.6,0}
\definecolor{codegray}{rgb}{0.5,0.5,0.5}
\definecolor{codered}{rgb}{0.8,0,0.0}
\definecolor{codeblue}{rgb}{0.0,0,0.8}
\lstdefinestyle{codeStyle}{
    backgroundcolor=\color{backcolour},   
    commentstyle=\color{codegreen},
    keywordstyle=\color{codeblue},
    numberstyle=\tiny\color{codegray},
    stringstyle=\color{codegray},
    numbers=left,                    
    tabsize=2
} 
\newcommand{\sfrac}[2]{#1/#2}

\makeindex

\begin{document}

\setboolean{onechapter}{true}

\pagestyle{fancy}
\lhead[\small\thepage]{\small\sf\nouppercase\rightmark}
\rhead[\small\sf\nouppercase\leftmark]{\small\thepage}
\newcommand{\innerfoot}{\footnotesize{\sf{\copyright} Feenstra \& Abeln}, 2014-2023}
\newcommand{\outerfoot}{\footnotesize \sf Intro Prot Struc Bioinf}
\lfoot[\outerfoot]{\innerfoot}
\cfoot{}
\rfoot[\innerfoot]{\outerfoot}
\renewcommand{\footrulewidth}{\headrulewidth}

\mainmatter
\setcounter{chapter}{13}
\chapterauthor{\HM*~\HMid \and \JvG~\JvGid \and \JG~\JGid \and \QH~\QHid \and \AM~\AMid \and \AR~\ARid \and \JV~\JVid \and 
\SA*~\SAid~~~\AF*~\AFid}
\chapterfootnote{* editorial responsability}
\chapterfigure{\includegraphics[width=0.5\linewidth]{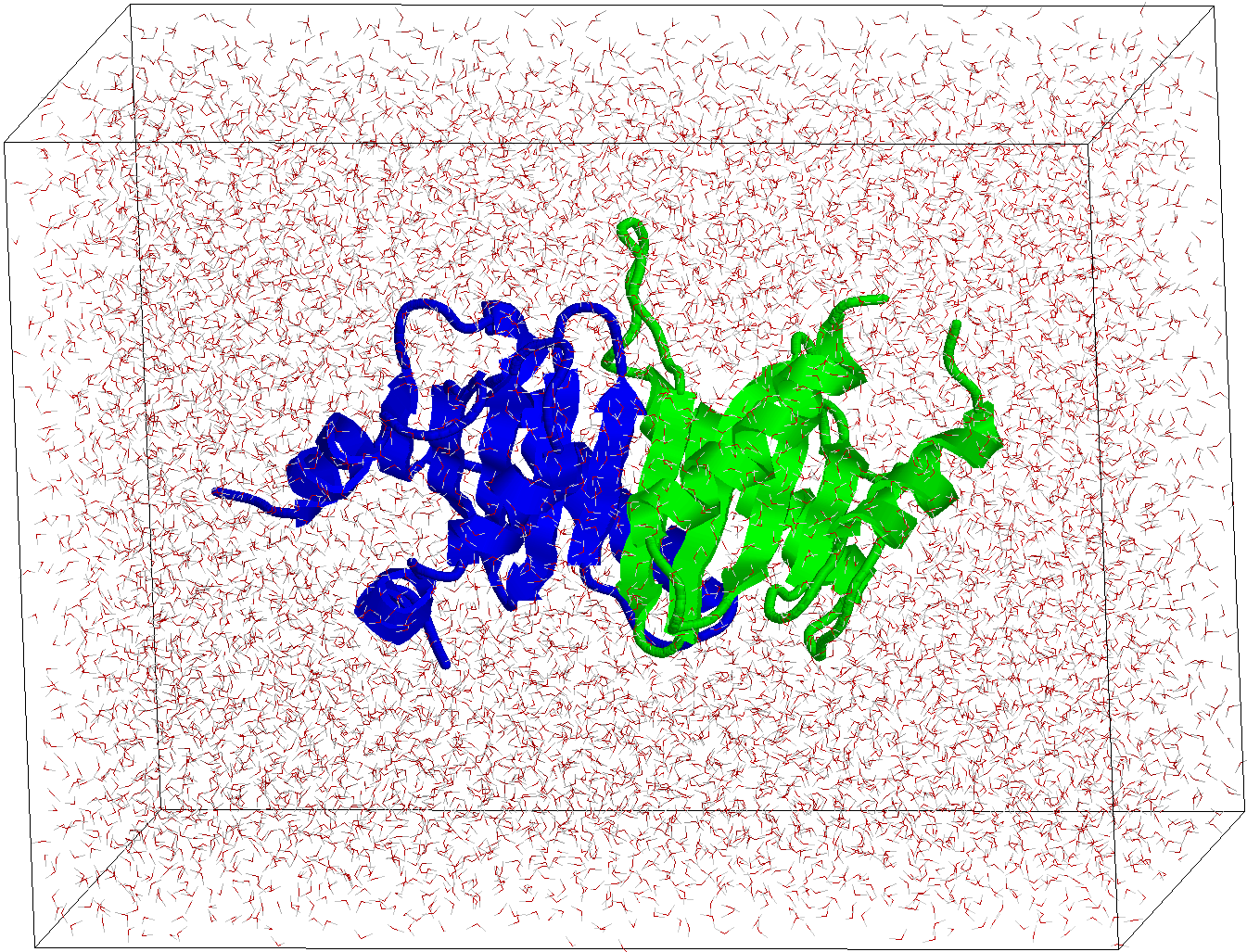}}
\chapter{Molecular Dynamics}
\label{ch:ChMD}

\ifthenelse{\boolean{onechapter}}{\tableofcontents\newpage}{} 

\section{Introduction to molecular dynamics} 
We know that many proteins have functional motions, and we already introduced this in \chref{ChDetVal}, \secref{ChDetVal:dyn}. One famous example presented there in \panelref{ChDetVal:allosteric}, is the cooperative binding of oxygen to hemoglobin, where the first oxygen binding induces a conformational change throughout the tetrameric structure, which make subsequent oxygen binding much more favourable. However, experimentally, such motions are hard to observe. Exceptions are when a protein can be crystallized in multiple distinct conformations, when fluorescence experiments are able to capture fast events, or in the very rare cases where time-resolved crystal structures can be recorded. A dramatic manifestation of functional motion occurs for some enzymes that are still active in the crystal form (which is relatively common). Large scale motions of the protein structure may occur during catalysis of the reaction, and sometimes these motions physically break the crystal when the substrate is added to the crystal! (This shows how strong molecular motions can be.)
As an alternative to experimentally studying protein motions, molecular dynamics (MD) simulations have been used since the 1960s. These allow us to simulate all atomic motions in detail, but of course within the restrictions of the accuracy of the molecular models used. 

In this chapter we will introduce MD simulations, which represent an important method to investigate the dynamic behaviour of proteins and polymers. Basically, in such simulations, the forces and interactions between particles are used to numerically derive the resulting three-dimensional movement of these particles over a certain time-scale. 
The main emphasis will lie on explaining the basic method of MD simulations, when and how to use them, and provide some examples of how to evaluate the results. More specifically, we will describe the physical interactions between atoms and the algorithms used to perform these simulations. Secondly, we will see how simulation results may be interpreted. For this purpose, some basic knowledge on thermodynamics and statistical mechanics, as introduced in \chref[nn]{ChThermo} is essential. It is particularly important to understand the relation between free energy and probability, in order to analyse simulation results.

Although a wide variation of problems (not necessarily of biological relevance) can be addressed using MD simulations, in the course of this book we will focus mainly on proteins.

\subsection{Simulating a protein by classical physics}

Molecular dynamics (MD) simulations, using classical or Newtonian physics, is currently the most common tool to investigate the dynamics and dynamic ensembles of proteins and other large biomolecules at a molecular level \citeeg{Adcock2006}. There are several branches of the technique that can be used to elucidate complicated or ambiguous results from biophysical experiments such as Atomic Force Microscopy (AFM) or Nuclear Magnetic Resonance (NMR) \cite{Kumar2010}. Most importantly, MD can provide insights into the molecular mechanisms through which proteins perform their functions; this can range from binding to a specific substrate, to proteins that can bind to each other, to the complete cycle of movement of molecular motors \citeeg{Karplus2005}.

Even though MD is used ubiquitously, it is important to remember that simulation methods and descriptions of interactions used are necessarily an abstraction of the real world (see in particular also the \panelref{ChMD:limits-newton}). Therefore, any hypothesis on mechanistic workings obtained from MD simulations should be verified with experimental evidence. This way, the biological, physical and scientific relevance of the results can be guaranteed.
Fortunately, Newtonian physics suffices for the vast majority of biologically relevant phenomena, because we are typically interested in macroscopic properties which depend on averages obtained from the ensembles observed in the simulation. Please refer to \chref{ChThermo} for more detail on the statistical dynamics of ensembles.

\begin{figure}[b]
\centerline{\includegraphics[width=0.6\linewidth]{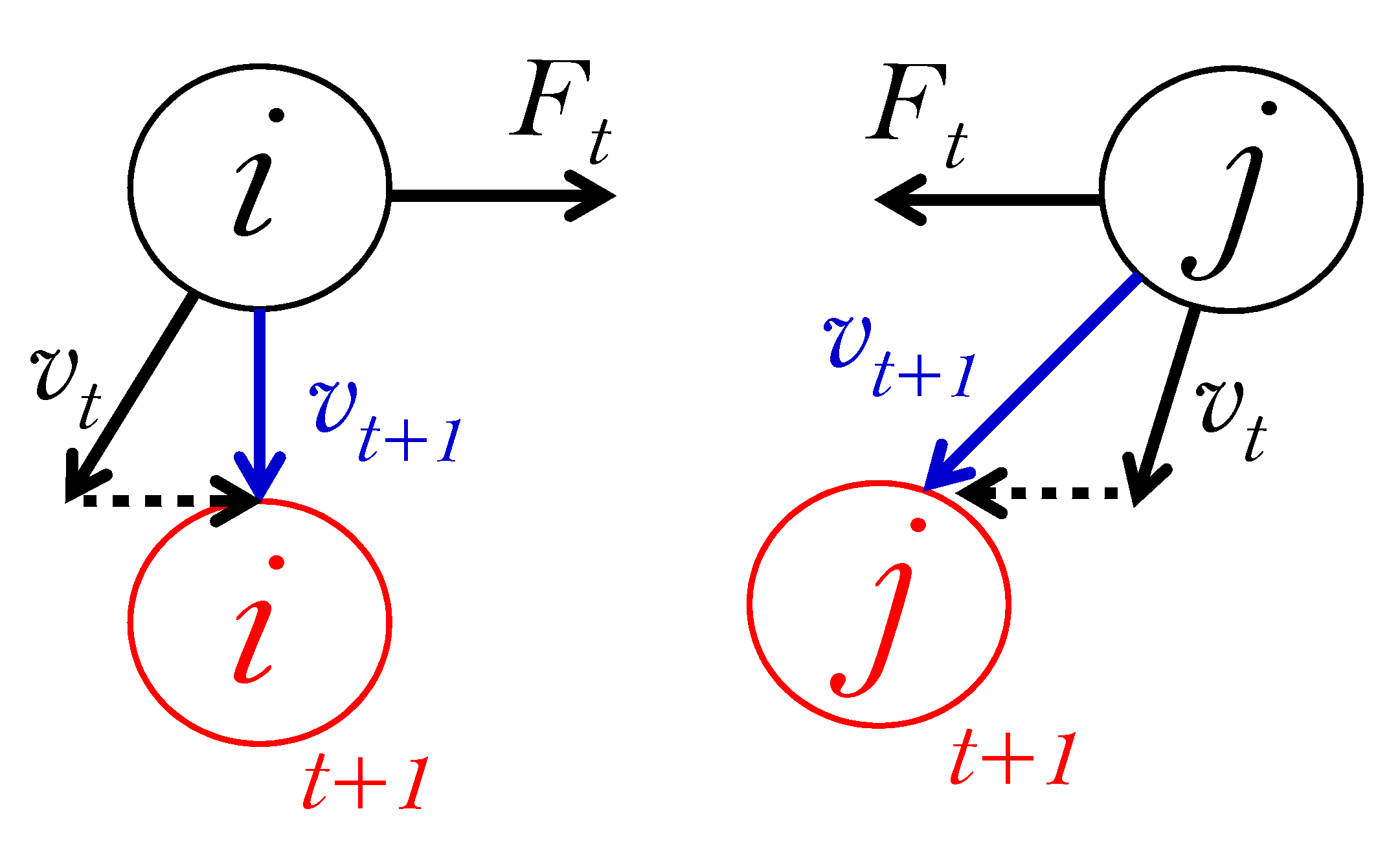}}
\caption{Two particles i and j at time t with initial position (indicated by black spheres) and velocities $v_t$ (black arrows), and exerting a force $F_t$ on each other (also black arrows; note that between two particles $F_{t;i,j} = - F_{t;j,i}$). These forces cause the velocities to change at the next time step $t + 1$, as indicated by $v_{t+1}$ in blue arrows, and the new velocities cause the positions to change as well (red spheres).}
\label{fig:ChMD-forces}
\end{figure}

In Newtonian physics, interactions are described as forces, and a force applied to an object triggers an effect. \figref{ChMD-forces} shows how the interaction (by way of force $F$) between two particles leads to changes in positions of both. This way, the structural and dynamical properties of large biological systems can be investigated at a molecular level. However, to simulate the large collection of atoms that make up biological systems, we need to resort to numerical approximations, as analytical solutions are not available (this is known as the `multi-body' problem in physics). Moreover, to obtain an accurate approximation, we need very small integration (time) steps, which is the main limit on computational efficiency. This is what much of the current chapter will deal with.

\section{Relevant time and length scales}

\begin{figure}
\centering
\includegraphics[width=0.8\linewidth]{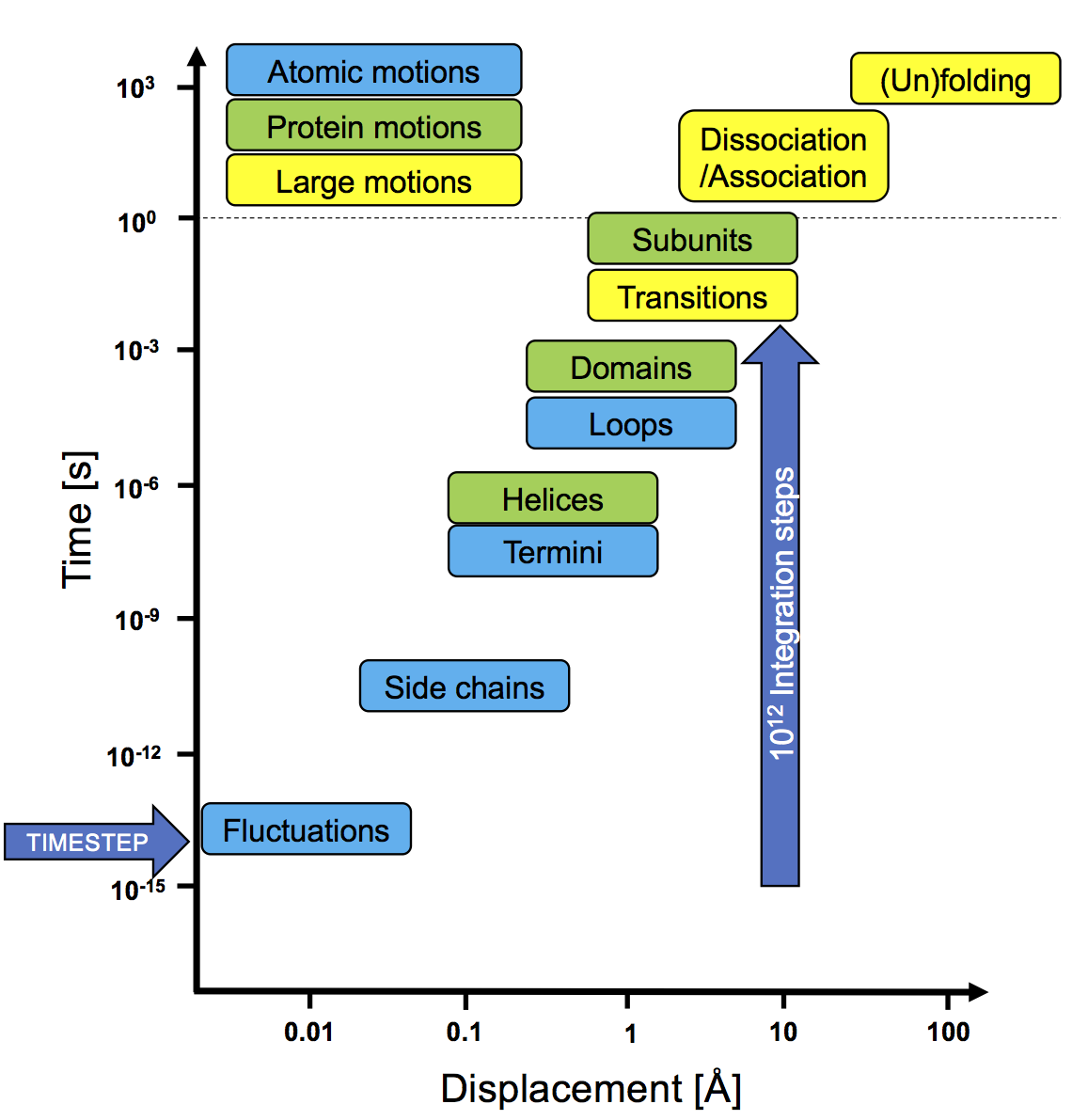}	
\caption{A simplified overview of the order-of-magnitude time and length scales of molecular motions that are of biological interest; the boxes indicate atomic motions (blue), protein domain motions (green), and large biological motions (yellow). The x axis corresponds to displacement of atoms or molecules, and the  y axis indicates the approximate time scales. Note that both axes are on a logarithmic scale. The size of the MD timestep (2 fs) is indicated by the arrow on the left; the arrow on the right indicates that a thousand billion ($10^{12}$) timesteps are needed to reach the shortest biologically relevant timescales.} 
\label{fig:ChMD:Timescales}
\end{figure}

To give some perspective to the phenomena we will be describing in molecular simulations, it is good to realize which time and length scales are involved. \figref{ChMD:Timescales} shows an overview of the relevant time and length scales of several classes of biological processes. Please note that these are order-of-magnitude indications that should not be taken as absolute values. Summarizing:
\begin{compactitem}
\item At the lower end of the scale are small-scale fluctuations of individual atoms over fractions of an {\AA}ngstrom (\AA, $10^{-10}$ m) distance and at timescales of tens to hundreds of femtoseconds ($10^{-15}$ s). 
\item Amino acid side chains that involve several to tens of atoms move tenths of an {\AA}ngstrom per picoseconds ($10^{-12} s$). 
\item Rotamer changes, conformational changes arising through rotation around rotatable bonds and sidechains, may produce larger motions, i.e., in the order of {\AA}ngstroms, and will be slower, tens to hundreds of picoseconds. 
\item The termini and loop regions, which involve several residues and thus many more atoms than a single sidechain are still slower, moving up to several Angstroms anywhere in the range of nano- ($10^{-9}$) and microseconds ($10^{-6}$). 
\item As helices are relatively stable structural (and dynamical) units, they move around the same scales as termini regions and loops. 
\item Larger parts, such as domains or subunits are again larger and therefore move slower, around tens of {\AA}ngstroms or even more and may take up to seconds in the slowest of cases. 
\item Conformational transitions take place in the same time-range as domain motives. These transitions often involve the motion of domain-size parts of a protein, or for example the association or dissociation of protein binding (e.g.\@ in protein-protein interactions). 
\item The typical folding time for a protein is on the order of seconds, nevertheless `fast folders' may take only milliseconds.
\end{compactitem}

For accurate simulation, the timestep used should be smaller than the fastest motions. Fastest vibrations involve hydrogen atoms (which are light), and/or particularly stiff angles, with vibrational periods of around 10 femtoseconds. Therefore, typically timesteps of 1 or 2 fs ($10^{-15}$s) are used, as indicated with an arrow at the bottom left in \figref{ChMD:Timescales}. It should be noted that an order of $10^{12}$ integration steps are then required to start getting into biologically relevant (millisecond) timescales as, indicated with a vertical arrow at the right in \figref{ChMD:Timescales}. This computational power is slowly coming within range of being feasible. But, in practice this means we are not routinely able to fold or unfold proteins in MD simulations, simply because we can not simulate for long enough. Note, also, that just simulating for longer may also not simply solve the problem, as the results would still depend on the accuracy of the force fields we use. We will deal with force fields in section \ref{ChMD:ff}. Moreover some effects, such as quantum effects, are hard to model accurately. In a large-scale survey of the ability of current force fields to improve upon homology models, \cite{Lindorff-Larsen2012,Raval2012} observed the accumulation of errors in long simulations, highlighting that this is an as yet unsolved problem.

Nevertheless, numerous interesting problems of biological relevance such as Amyloid aggregation in Alzheimer's \cite{Lemkul2013} and T-cell receptor-MHC interactions \cite{Cuendet2011} have been investigated using simulations that might strictly be considered too short. These simulations have fortunately shown the interesting and relevant dynamical molecular behaviour that helps answering several important biological questions. 

\begin{bgreading}[Historical background]
A historical overview on the evolution and applications of MD simulations is given in \cite{Gunsteren2006}. The rough estimates of future simulation times are extrapolated from past achievements based on the size of those systems and assuming the continued growth of computer power by Moore's law by doubling every 18 months \cite{Moore1965}. Assuming that will hold, it may take till the end of this century before we can simulate protein folding at the speed it occurs in nature -- but even then only a single protein, and only if the accuracy of our force fields do not remain limiting. 

\centerline{\small
\newcommand{\smct}[1]{{\footnotesize{\citeauthor{#1}}}}
\begin{tabular}{@{}l>{\raggedright\arraybackslash}p{0.45\linewidth}@{}c@{}c@{}}
\hline
year & system & time-scale & reference \\
\hline
1936   & Gelatine balls              &       & \smct{Morrell1936} \\
1953   & MC Simulations              &       & \smct{Metropolis1953} \\
1957   & MC of Lennard-Jones spheres &       & \smct{Wood1957} \\
1964   & MD of liquid Argon          & 10 ps & \smct{Rahman1964} \\
1970's & Non-equilibrium methods     &       & \\
1970's & Stochastic dynamics methods &       & \\
1974   & MD of liquid water          &       & \smct{Stillinger1974} \\
1977   & MD of protein in vacuum     & 20 ps & \smct{McCammon1976} \\
1980's & Quantum-mechanical effects  &       & \\
1983   & MD of protein in water      & 20 ps & \smct{Gunsteren1983} \\
1998   & MD of reversible peptide folding & 100 ns   & \smct{Daura1998}\\
1998   & MD of protein folding       & 1 $\mu$s & \smct{Duan1998} \\
2010   & MD of reversible small protein folding & 1 ms & \smct{Shaw2010}\\
\hline\hline
Today & Large proteins or complexes in water or membrane & 
 \multicolumn{2}{>{\raggedright\arraybackslash}p{0.4\linewidth}}{up to milliseconds ($\sim 10^{12}$ -- $10^{14}$ slower than nature)} 
                                      \\
\hline\hline
2029 &  Protein folding             &  1 ms                & ??? \\
2034 & E-coli, $\sim 10^{11}$ atoms  & 1 ns                 & \\ 
2056 & Eukaryotic cell, $\sim 10^{15}$ atoms & 1 ns         & \\
2080 & Protein folding            & \multicolumn{2}{l}{as fast as in nature} \\
\hline
\end{tabular}
}\label{tab:ChMD:History}
~\\

The field of molecular simulations was initiated around 1930, when Morrell and Hildebrand \cite{Morrell1936} investigated the distribution of gelatine spheres, compared to X-ray experiments on atoms. The subsequent rapidly growing number of applications was enabled through to the continuous development and improvement of computational facilities over the last 50 years. This allowed for more and more detailed theoretical investigations of larger systems over longer timescales. Along with this development, simulations have moved from hard-sphere systems \cite{Alder1957} and simple mono-atomic systems (e.g.\@ argon) in the 60's \cite{Rahman1964}, to proteins and water separately in the 70's \cite{Stillinger1974,McCammon1976}, to proteins in water in the 80's \cite{Gunsteren1983}, and finally, to something close to the equilibrium behaviour of peptides and (small) proteins in water in the 90's and early 2000's \cite{Daura1998,Duan1998,Shaw2010}. Generally it can be said that MD simulation became important for biophysics in the late 70's and early 80's with the first simulations of a protein in water. More biological relevance perhaps can be said to start with the simulation of reversible, equilibrium peptide folding (late 90's) and protein folding (early 2000's).
\end{bgreading}

\section{Forces \& interactions} 
If we want to understand how particles move, we need to understand how they interact with each other, or more precisely, with what forces they repel or attract each other. The basis of all molecular simulations is the so-called \textit{force field}. Force fields consist of a collection of all forces which are considered to occur in a system of interest, for example all atoms within a protein and the surrounding solvent. Forces, as noted before, correspond to interactions, in this case between individual atoms in the system. Note that these interactions and therefore the energies associated with them are dependent on the types of atoms involved, as well as their positions with respect to each other. Formally, the force $F$ on particle $i$ is the derivative of the energy $U$, depending on the position $r_i$ of particle $i$, and is given as:

\begin{equation}
        F_i = - \frac{\partial U}{\partial r_i},
\end{equation}

\subsection{Force fields} 
\label{ChMD:ff}

Force fields can be written as interaction energies; in the next section on \secref[NN]{ChMD:interactions} we will see how interaction forces and interaction energies are related. In molecular mechanics,  interaction energies between individual atoms in the system contain the following terms for interaction energies:
\begin{equation}
U_{total} = U_{bonded} + U_{non-bonded} + U_{crossterm}
\label{eq:ChMD:energies}
\end{equation}
Thus, in molecular systems, a distinction is made between bonded interactions, non-bonded interactions and `other' interactions called `crossterm' in the formula. Bonded interactions act in between atoms close by in the same molecules; most notably two atoms in a bond, three in an angle and four in a dihedral: 
\begin{equation}
U_{bonded} = U_{bond} + U_{angle} + U_{dihedral}
\label{eq:ChMD:bondeds}
\end{equation}
Non-bonded interactions act in between molecules and also in between atoms in the same molecule: these are Coulomb's (electrostatic) and Van der Waals' (atomic contact) interactions. 
\begin{equation}
U_{non-bonded} = U_{Coulomb} + U_{Van der Waals}
\label{eq:ChMD:nonbondeds}
\end{equation}
Combining \eqref[s]{ChMD:energies}-\ref{eq:ChMD:nonbondeds}, the total energy may thus be specified as:
\begin{equation}
U_{total} = U_{bond} + U_{angle} + U_{dihedral} + U_{Coulomb} + U_{Van der Waals} + U_{crossterm} . 
\label{eq:ChMD:energy}
\end{equation}

Each of these energy terms will be specified in more detail in the next section. The main assumption here is that the total interaction (e.g.\@ between molecules) can be described as a sum of pairwise atomic interactions. For some specific interactions, e.g., weak non-covalent interactions between two different bonds, a pairwise description is not a good approximation; these are represented as the `crossterm' interactions in the formula. For protein simulations these terms are, usually, negligible and therefore not included in the most commonly used force fields to save computational resources. 

\subsection{Interactions}
\label{sec:ChMD:interactions}

\begin{figure}
\centerline{
\includegraphics[width=0.8\linewidth]{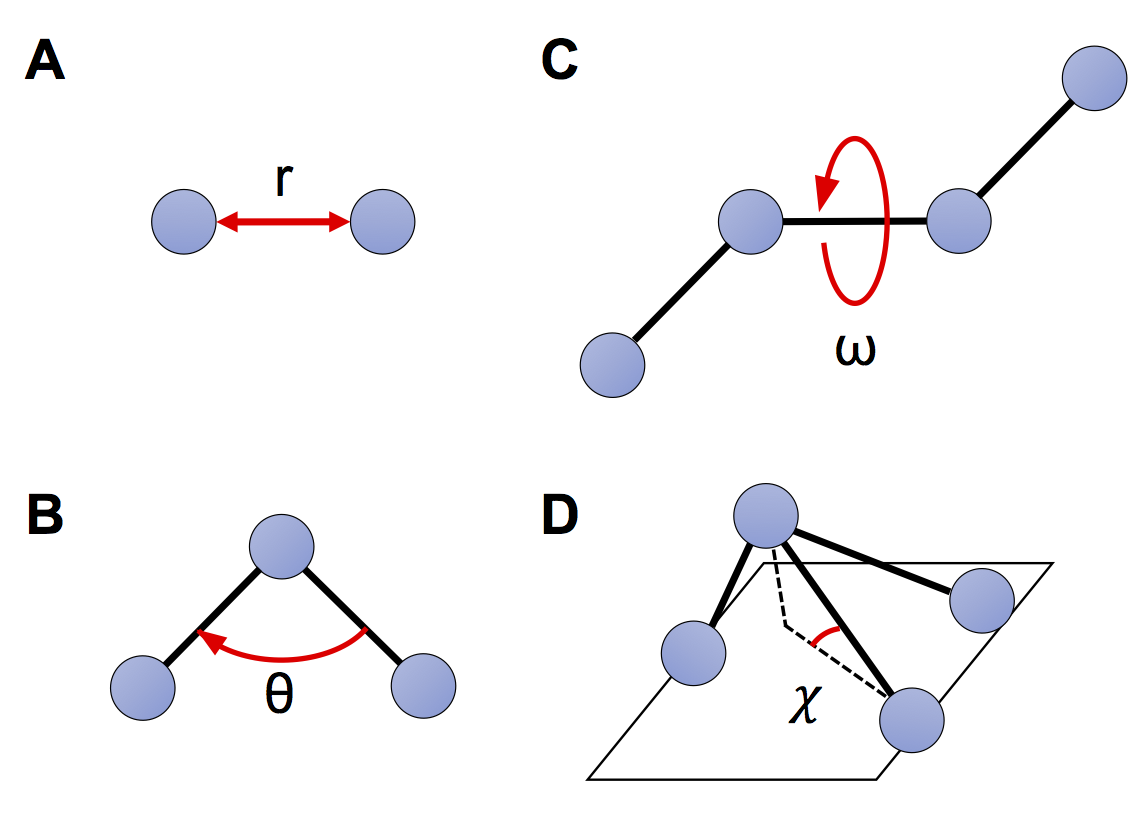}}
\caption{Schematic of common bonded interactions: \textbf{A} bond between two atoms, \textbf{B} angle between three atoms, \textbf{C} dihedral angle between four atoms (rotatable bond between the middle two), \textbf{D} improper dihedral, used to fix particular orientation, e.g.\@ in-plane, or as in the drawing one atom out of the plane of three other atoms. 
\label{fig:ChMD-bondeds}}
\end{figure}

The most commonly used energy functions for the bonded interactions are bond, angle, and dihedral; for the non-bonded these are Coulomb and Van der Waals. Bond and angle interactions are shown in \figref[a]{ChMD-bondeds} and \hyperref[fig:ChMD-bondeds]{b}. Proper dihedrals correspond to rotatable bonds, for the proteins backbone these are the N-C$_\alpha$ and C$_\alpha$-C=O bonds, as well as the non-aromatic ones in the side chains, \figref[c]{ChMD-bondeds}. Improper dihedrals are used to constrain a fixed geometry, as for example the peptide bond or the tetrahedral arrangements of the hydrogens in an -NH$_3$ group, see \figref[d]{ChMD-bondeds}. For bonds, angles and improper dihedrals, typically quadratic (harmonic) functions are used:
\newcommand{\myphantom}{%
  \phantom{\textrm{Improper dihedrals:} ~} &%
  \phantom{~ U(r_{i j})=4\epsilon[(\sigma/r_{i j} )^{12}-(\sigma/r_{i j})^6]} %
  \\[-\baselineskip]}
\begin{equation}\label{eq:ChMD:bonds} 
  \begin{aligned}\myphantom
  \textrm{Bonds:} ~&~ U(r ) = \tfrac12 k_b (r-r_0)^2    
  \end{aligned}
\end{equation}
\begin{equation}\label{eq:ChMD:angles} 
  \begin{aligned}\myphantom
  \textrm{Angles:} ~&~ U(\theta) = \tfrac12 k_\theta (\theta-\theta_0)^2 
  \end{aligned}
\end{equation}
\begin{equation}\label{eq:ChMD:idihs} 
  \begin{aligned}\myphantom
  \textrm{Improper dihedrals:} ~&~ U(\chi) = \tfrac12 k_\chi (\chi-\chi_0)^2 
  \end{aligned}
\end{equation}
\noindent
For these interactions, $r$ denotes distance, and $\theta$ (theta), and $\chi$ (chi) denote angles. The subscript $0$ refers to the equilibrium value in each case. The $k$'s are force constants. These parameters depend on the atoms involved.

(Proper) dihedrals are expressed as (periodic) cosines:
\begin{equation}\label{eq:ChMD:dihs} 
  \begin{aligned}\myphantom
  \textrm{Dihedrals:} ~&~ U(\omega) = \tfrac12 \sum k_j  [1 + (-1)^{ j+1} cos ( j \omega + \phi ) ]
  \end{aligned}
\end{equation}
Here, $\omega$ (omega) denotes the angle.
Note that angle (\eqref{ChMD:angles}) and dihedral (\eqref[s]{ChMD:idihs} and \ref{eq:ChMD:dihs}) bonded interactions are not strictly speaking pairwise as they involve three and four atoms, respectively. They are the main non-pairwise contribution that is usually included for protein simulations.

\begin{figure}
\centerline{\includegraphics[width=0.8\linewidth]{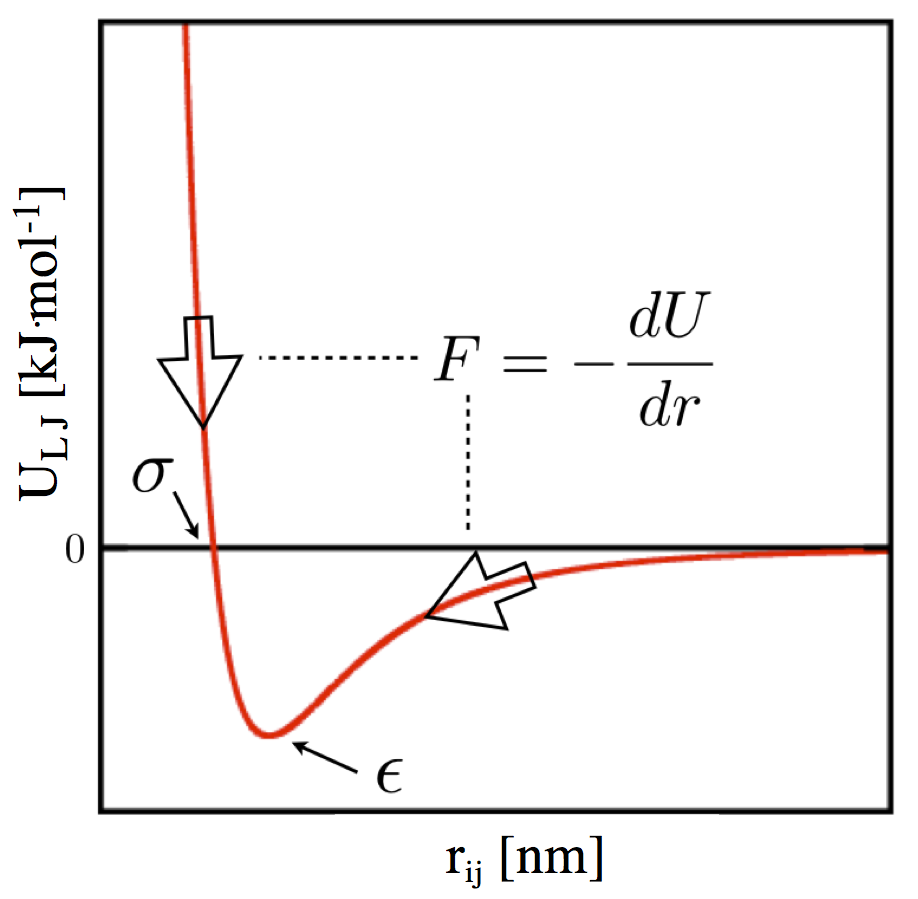}}
\caption{The Lennard-Jones potential $U_{LJ}$ as function of $r_{ij}$. The force $F$ is the derivative of $U$ with respect to $r$, or in other words the slope of the function $U$ in this plot (indicated by the two arrows).
}
\label{fig:ChMD-lennard-jones}
\end{figure}

Non-bonded interactions are modelled according to Coulomb's law for electrostatic forces, and the Lennard-Jones potential describing Van der Waals forces, respectively as:
\begin{equation}\label{eq:ChMD:coul} 
  \begin{aligned}\myphantom
  \textrm{Coulomb:} ~&~ U(r_{i j} ) =  \frac{\epsilon_0 ( q_i \cdot q_j ) }{ r_{i j} }
  \end{aligned}
\end{equation}
\begin{equation}\label{eq:ChMD:LJ}
  \begin{aligned}\myphantom
  \textrm{Lennard-Jones:} ~&~ 
    U(r_{i j} ) = 4 \epsilon \left[ %
           \left(\frac{\sigma}{r_{i j}}\right)^{12} - %
           \left(\frac{\sigma}{r_{i j}} \right)^6 \right]
  \end{aligned}
\end{equation}
For the non-bonded interactions, $r_{i,j}$ denotes the distance between atoms $i$ and $j$, and $q_i$ and $q_j$ the charges on atoms $i$ and $j$. In \eqref{ChMD:coul} $\epsilon_0$ (epsilon) indicates Coulomb's constant and in \eqref{ChMD:LJ} $\sigma$ (sigma) and $\epsilon$ indicate the Van der Waals radius and the strength of the interaction, respectively. These parameters depend on the atoms involved.  The Lennard-Jones potential is a combination of repulsive interactions originating from the Pauli exclusion principle, and attractive interactions resulting from London dispersion forces. \figref{ChMD-lennard-jones} shows an example of a Lennard-Jones interaction potential. 
Generally speaking, an energy is associated with the forces of the system: the slope of the energy determines the direction in which the atom is pushed as shown in \figref{ChMD-lennard-jones}. 

\subsection{Parameters}
The parameters included in force fields are usually derived from experimental techniques and (expensive) quantum chemical calculations \cite{Leach}. For example, bond lengths and angles, i.e.\@ $r_0$, $\theta_0$, $\chi_0$, as introduced in the previous section \secref[N]{ChMD:interactions}, are based on distances in small molecule crystal structures, while force constants are based on a combination of infrared spectroscopy (which measures molecular vibrations) and quantum chemical calculations. Existing force fields are continuously validated and improved to properly describe the nature of a given system.  Many modern force fields like GROMOS \cite{Gunsteren1996,Oostenbrink2004}, AMBER \cite{Case2014}, CHARMM \cite{Brooks1983} and OPLS \cite{Kaminski2001} are extensively validated on thermodynamic parameters like partition coefficients (affinity for amino acids side chains for either water or oil phase), and peptide and protein folding equilibria \cite{Daura1999,Gunsteren2001,Oostenbrink2004,Case2014,Brooks1983, Shaw2010,Lindorff-Larsen2012}. This makes these force fields ready to simulate biological phenomena, since we are usually interested in the equilibrium states and transition events which are thermodynamic quantities. See \chref{ChThermo} for more on thermodynamics.

\begin{bgreading}[Limitations of Newtonian physics and force fields]
\label{panel:ChMD:limits-newton}
MD simulations take a classical mechanical approach by using Newton's second law of motion to determine the collective dynamics of a set of particles, such as the atoms in a protein and the surrounding solvent (typically water); this is known as the Molecular Mechanics (MM) approach. Most simulations are performed without considering quantum effects explicitly, such as pi-pi interactions between aromatic rings (e.g.\@ Phenylalanine), electron transfer, proton transfer, bond breakage, and even H-bonds in certain conditions. 

For some chemical processes such as enzyme catalysis, proton transfer, and pH effects (typically involving hydrogen atoms) simulations using more advanced models that incorporate quantum physics may be required. Unfortunately, when compared to a classical mechanical approach, quantum mechanical (QM) methods require much longer computation times to investigate the conformational space and energy landscape for large systems such as proteins and polymers, which typically consist of many tens of thousands of atoms. QM methods are currently limited to hundreds of atoms at the very best. Hybrid QM/MM approaches may also be employed, but these also remain strongly limited by the expense on the QM part.

Fortunately, many biological effects that we may be interested in are macroscopic properties, such as protein folding or ligand binding, for which classical or Newtonian physics suffices. Moreover, for such macroscopic properties, our models can be, and typically also have been, validated with experiments.\cite{Gunsteren2006} 

However, you have to keep in mind that force fields cannot automatically adapt to new conditions, which means that they will work only for atoms and structures for which parameters have been appropriately determined. If your system includes anything not described in the force field, e.g., a ligand or metal ions, this will typically make the software crash. Therefore, you need to parametrize your ligands and metal coordination centers before launching such a simulation. This is not trivial although several approaches and web tools are available to generate reasonably good topologies for small organic molecules. One of the major bottlenecks is that classical force fields describe electrostatic interactions in physical systems as static atom-centered charges that do not change during your simulation. In other words, the polarization that a real physical system undergoes in a dielectric medium such as water is entirely neglected. To address this problem, polarizable force fields that aim to describe the variations in charge distribution within the dielectric environment a very promising for future applications \cite{Halgren2001_PFF}.
However, these force fields are not easy to set up and therefore currently not available for large systems. 

If you need to include metal ions (for example in metalloproteins), things may even get more complicated. Here, transition metals such as copper,  zinc and iron are particularly challenging as they exhibit several oxidation states. Often, a subtle equilibrium between two coexisting redox states will be present in a biological systems. Until now, even when using hybrid methods such as QM/MM, it is not possible to get sufficiently realistic parameters for metalloprotein simulations. However, there exist several approaches that approximate these systems as best as possible. For the interested reader, the review by \citet{Merz_2017_metals} provides a detailed overview of the advances, state-of-the-art and bottlenecks of modelling metal ions in classical dynamics.
\end{bgreading}

Now, because of the assumption of additive pairwise interactions, you can combine everything (\eqref[s]{ChMD:bonds}-\ref{eq:ChMD:LJ}) by simply adding all energy terms, as was already suggested by \eqref{ChMD:energy}, which leads to the following result:
\begin{equation}
\begin{aligned}
U(r) = & \tfrac12  \sum_b k_b (r-r_0)^2 + \tfrac12 \sum_\theta k_\theta (\theta-\theta_0)^2 + \\
& \tfrac12 \sum_\omega \sum k_j  [1 + (-1)^{ j+1} cos ( j \omega + \phi ) ] + \\
& \tfrac12 \sum_\chi k_\chi (\chi-\chi_0)^2 + \epsilon \sum_i \sum_j ( q_i \cdot q_j ) / r_{i j} + \\
& 4 \epsilon \sum_i \sum_j [ (\sigma/r_{i j} )^{12} - (\sigma/r_{i j} )^6 ]
\end{aligned}
\end{equation}

Although not all details of this formula are equally important, you should notice the character of the different summations. The first four are single sums over all bonds ($b$), angles ($\theta$) and proper ($\omega$) and improper ($\chi$) dihedrals. Here, as above, we have simply left out cross terms between bonds, angles or dihedrals; this is the `other' class already mentioned which is typically negligible for protein simulations. The lasts two sums go over all unique pairs of atoms $ij$ for both the Van der Waals (first) and Coulomb (second) interactions. The number of bonds, angles and dihedrals in a system increases approximately linearly with the number of atoms. However, the pairs of atoms is quadratic in the number of atoms. Therefore, the non-bonded interactions are typically the vast majority ($>95\%$) of the computational effort of any biomolecular simulation \cite{Leach}.

\begin{table}
\caption{Overview of important molecular dynamics force fields.}
\centerline{
\begin{tabular}{>{\raggedright\arraybackslash}p{0.2\linewidth}>{\raggedright\arraybackslash}p{0.5\linewidth}>{\raggedright\arraybackslash}p{0.35\linewidth}}
\hline
\bf Force fields & \bf Description & \bf Ref.\\
\hline
AMBER & Full-atomistic, used for proteins and DNA & \citealt{Ponder2003,Tian2020} \\
CHARMM & Full-atomistic, used for small molecules and macromolecules & \citealt{Brooks1983,BrooksBrooks2009} \\
GROMOS & Full-atomistic, bio-molecular systems: solutions of proteins, nucleotides, and sugars & \citealt{Oostenbrink2004,Reif2012} \\
 OPLS & Full-atomistic, Optimized Potential for Liquid Simulations & \citealt{Kaminski2001,Shivakumar2012} \\
MARTINI & Coarse grained, molecular dynamics simulations of large bio-molecular systems ($\sim$4 heavy atoms represented by a single bead) & \citealt{Marrink2007,Monticelli2008,Souza2021} \\
\hline
\end{tabular}
}
\footnotesize Note that force fields are constantly improved, so that there are several versions available for each. Therefore, it is recommended to use the latest version of a force field (unless you have a valid reason not to). The choice of the force field itself is not strictly predetermined and may often depend on the system of interest, the program package or the history and experience of the research team. A relatively recent review of force fields can be found in \citealt{Lindorff-Larsen2012}.
\end{table}

\begin{table}
\caption{Overview of important molecular dynamics software packages.}
\newlength{\plength}
\setlength{\plength}{0.3\linewidth}
\newcommand{\pcell}[2][t]{\parbox[#1]{\plength}{\raggedright #2}}
\centerline{
\begin{tabular}{>{\raggedright\arraybackslash}p{\plength}>{\raggedright\arraybackslash}p{0.5\linewidth}>{\raggedright\arraybackslash}p{0.25\linewidth}}
\hline
\bf Software packages & \bf Description & \bf Ref. \\
\hline
\pcell{\textbf{AMBER} -- 
Assisted Model Building with Energy Refinement} & Molecular dynamics simulations of bio-molecules & \citealt{Case2014} \\
\pcell{\textbf{GROMACS} -- 
GROningen MAchine for Chemical Simulations} & Molecular dynamics simulations of proteins, lipids and nucleic acids & \citealt{Pronk2013,Abraham2015} \\
\pcell{\textbf{NAMD} --
NAnoscale Molecular 
Dynamics program} & Molecular dynamics simulations of large bio-molecular systems & \citealt{Phillips2005,Phillips2020} \\
\pcell{\textbf{YASARA} --
Yet Another Scientific Artificial Reality Application} & Molecular dynamics program, including molecular visualization and modeling & \citealt{Krieger2014,Krieger2015} \\
\pcell{\textbf{CP2K} --
Open Source molecular Dynamics} & Atomistic and molecular simulations of solid state, liquid, molecular, and biological systems including QM/MM 
& \citealt{Warshel1976} \\
\hline
\end{tabular}
}
\end{table}

\section{Dynamics}
Now that we have the forces (and energies) of our system, we can look at the mechanics behind the simulations. We will first look into dynamics in the form of MD simulations. In the next chapter (\chref[N]{ChMC}) we will cover Monte Carlo (MC) simulations, which can be seen as a statistical approach to the same problem that MD simulations address. 

To describe a dynamic system, the coordinates (positions, $r$) and momenta (velocities times the mass, $p = m \cdot v$) of all particles are needed. These two sets of variables describe the `phase space'. Previously, in \chref{ChThermo} we have already introduced the \emph{conformational space} which only includes the coordinates (and not the velocities or momenta). In the course of this chapter, we will not go into the subtle difference between these, so for simplicity we will consider the conformational space an adequate description of our molecular system.

In addition to the coordinates and momenta, the energies of the particles are needed. These are split into two parts: the potential and the kinetic energies, $U$ and $K$ respectively. The potential energy derives from the force field description we introduced in the previous section, while the kinetic energy simply derives from the temperature of the system. For example at $0~K$ (in a frozen system) all velocities are zero, whereas at higher temperature the (average) velocity of all atoms increases.

The final ingredient required to understand the course of MD simulations is the relation between the forces (derivative of the potential energy) and the positions of the atoms, which goes according to Newton's famous law of motion: 
\begin{equation}
        F = m \cdot a .
\end{equation}
Here, $F$ is the force, $a$ the acceleration and $m$ the mass. As we can primarily calculate the change in velocity, this equation can be rearranged to 
\begin{equation}
        a = F/m .
\end{equation}
Actually, Newton wrote his law as a differential equation 
\begin{equation}
        \frac{\partial a(r)}{\partial r} = \frac1m ~ \frac{\partial F(r)}{\partial r},
\end{equation}
with $r$ the position, $F$ and $a$ as before, and $\partial$ indicating the (partial) first derivative. Remember that the differential of the energies is the force, so for dynamics we require a differentiable function for the energy. 

Now, solving Newton's equation of motion yields a description of the motion of the particles (atoms) involved, which is what we want when we perform an MD simulation. However, this differential equation can only be solved analytically for a system of two particles. Therefore, in a typical MD simulation dealing with large number of particles, a numerical approximation is required to obtain the changes in velocity and position over time. The analytical solution to Newton's equation has nice properties: it is energy-conserving, reversible and deterministic. The numerical approximation, however, may not be so well behaved. Moreover, for a system of more than two particles, we know that this behaves intrinsically non-deterministic or `chaotic'; meaning that even the tiniest differences in the starting situation will eventually diverge into vastly different behaviour of the simulation. In practice however, for our biomolecular simulations, this is not a problem as average and/or aggregate behaviour should be independent of the details of the integration scheme (as long as the errors are not systematic). 

\subsection{Integrating equations of motion}

\subsubsection{The Verlet integration scheme}

\begin{figure}
\begin{lstlisting}[language=Python,mathescape]
def molecular_dyanamics(r,num_steps,N,V,T):
    # generate initial velocities, 
    # depending on set Temperature
    v = initial_velocities(T)
    for i in range(num_steps):
        # compute forces, from current coordinates r; 
        # depends on (globally defined) force field
        F = calc_forces(r)
        # calculate thermostat scaling factor for set 
        # temperature (depends on velocities)
        $\lambda$ = temp_scaling(v)
        # calculate barostat scaling factor for set 
        # pressure (depends on coordinates and forces)
        $\mu$ = pressure_scaling(r, F)
        # using calc'd forces F, update velocities v
        v$'$ = $\lambda$*(v + F/m*$\Delta$t)
        # using calculated velocities v, 
        # update coordinatess r
        r$'$ = r + v$'\cdot\Delta$t
        # apply constraints - certain atomic distances 
        # and angles that are not allowed to change, 
        # like for hydrogen atoms
        r$''$ = apply_constraints(r$'$, r)
        # correct v for constraints
        v = (r$''$ - r)/$\Delta$t
        # apply barostat and obtain final coordinates
        r = $\mu$*r$''$
        # every set number of steps, perform output
        # of system variables, r, v, or F
        if (i%nst_log) do_print_log(r, v)
        if (i%nst_r) do_output_r(r)
        if (i%nst_v) do_output_v(v)
        if (i%nst_F) do_output_F(F)
    # end for loop
# end Molecular Dynamics
\end{lstlisting}
\caption{Molecular Dynamics algorithm for molecular simulations in pseudo code Python style.}
\label{fig:ChMD:MDpseudo}
\end{figure}

The most used numerical scheme to integrate the equations of motion in MD simulations is the 'Verlet integration' scheme, which is based on a Taylor expansion of Newton's differential equation \cite{Verlet1967}. Using a Taylor expansion is one way of writing a differential equation, which uses infinitesimally small increments, as a numerical approximation using finite size time step increments (See \panelref{ChMD:Verlet} for more details). Note that there exist a variety of additional molecular dynamics integrators that can be used. However, a detailed discussion of the differences between the methods, their benefits and limitations goes beyond the scope of this chapter and further details may easily be found elsewhere (see for instance \citet{Leach} or \citet{FrenkelSmit}). 

\begin{panel}[Derivation of the Verlet integration scheme]
\label{panel:ChMD:Verlet}
The trick to deriving the Verlet integration scheme is to take the Taylor expansion of Newton's equation of motion for the forward $(t+\Delta t)$ and backward $(t-\Delta t)$ step in time. Adding these cancels the third order term (eliminating any approximations there), and yields a formula of positions at the next time point $(t+\Delta t)$ as a function of the current $(t)$ and previous $(t-\Delta t)$) time points, which contains only zero, second and fourth order terms. If our final equations include only the zeroth and second order terms, we thus make only a fourth order error. 
\begin{equation}
{\bf r}(t+\Delta t) = {\bf r}(t) + {\bf r'}(t)\Delta t + \tfrac{1}{2!}{\bf r''}(t)\Delta t^2 + \tfrac{1}{3!}{\bf r'''}(t)\Delta t^3 + \dots 
\end{equation}
where the first derivative (the velocity) is ${\bf r'}(t) = {\bf v}(t)$ and the second derivative (the acceleration) is ${\bf r''}(t) = {\bf a}(t)$.  
\begin{equation}
\begin{aligned}
{\bf r}(t+\Delta t) & = {\bf r}(t) & +~{\bf r'}(t)\Delta t & +~\tfrac{1}{2!}{\bf r''}(t)\Delta t^2 & +~\tfrac{1}{3!}{\bf r'''}(t)\Delta t^3 \\
{\bf r}(t-\Delta t) & = {\bf r}(t) & -~{\bf r'}(t)\Delta t & +~\tfrac{1}{2!}{\bf r''}(t)\Delta t^2 & -~\tfrac{1}{3!}{\bf r'''}(t)\Delta t^3 \\ \hline 
& & & & + \\[-2ex]
{\bf r}(t+\Delta t) + {\bf r}(t-\Delta t) & = 2{\bf r}(t) & & + 2\tfrac{1}{2!}{\bf r''}(t)\Delta t^2 \\
\end{aligned}
\end{equation}

Which, using $2\tfrac{1}{2!}=1$ and ${\bf r''}(t)={\bf a}(t)$  can then be rearranged to:
\begin{equation}
{\bf r}(t+\Delta t) = 2{\bf r}(t) - {\bf r}(t-\Delta t) + {\bf a}(t)\Delta t^2 
\end{equation}

One can further rewrite this using the velocity ${\bf v}(t-\tfrac12\Delta t) = {\bf r}(t) - {\bf r}(t-\Delta t)$ into:
\begin{equation}
{\bf r}(t+\Delta t) = {\bf r}(t) + {\bf v}(t-\tfrac12\Delta t) + {\bf a}(t)\Delta t^2 
\end{equation}
\end{panel}

\begin{panel}[Temperature and pressure]
\label{panel:ChMD:Temp}
The integration schemes used in MD simulations conserve energy, and the simulations are performed in a box of fixed dimensions. Thermodynamically speaking, these are thus in the $NVE$ ensemble, where the number of particles ($N$, atoms), volume ($V$) and energy ($E$) are constant. As noted in \chref{ChThermo}, most biological experiments, however, are done in the so-called `grand canonical ensemble' where chemical potential ($\mu$), Pressure and Temperature are constant. This is very hard to achieve in general in a simulation \cite{Pool2012}, so as the next best option, a typical MD simulation is set up in the $NPT$ ensemble where, as before, the number of particles is constant, and pressure ($P$) and temperature ($T$) are kept constant on average over some relatively short time period. There are simple solutions for maintaining constant temperature and pressure, called `weak coupling', that have been used for many decades \cite{Berendsen1984}, but currently more recent methods are generally used as they avoid some serious problems in long time-scale simulations \cite{ChengMerz1996,Mor2008,Lingenheil2008}.
\end{panel}

\subsubsection{Choosing timesteps}
\begin{figure}
\centerline{\includegraphics[width=0.8\linewidth]{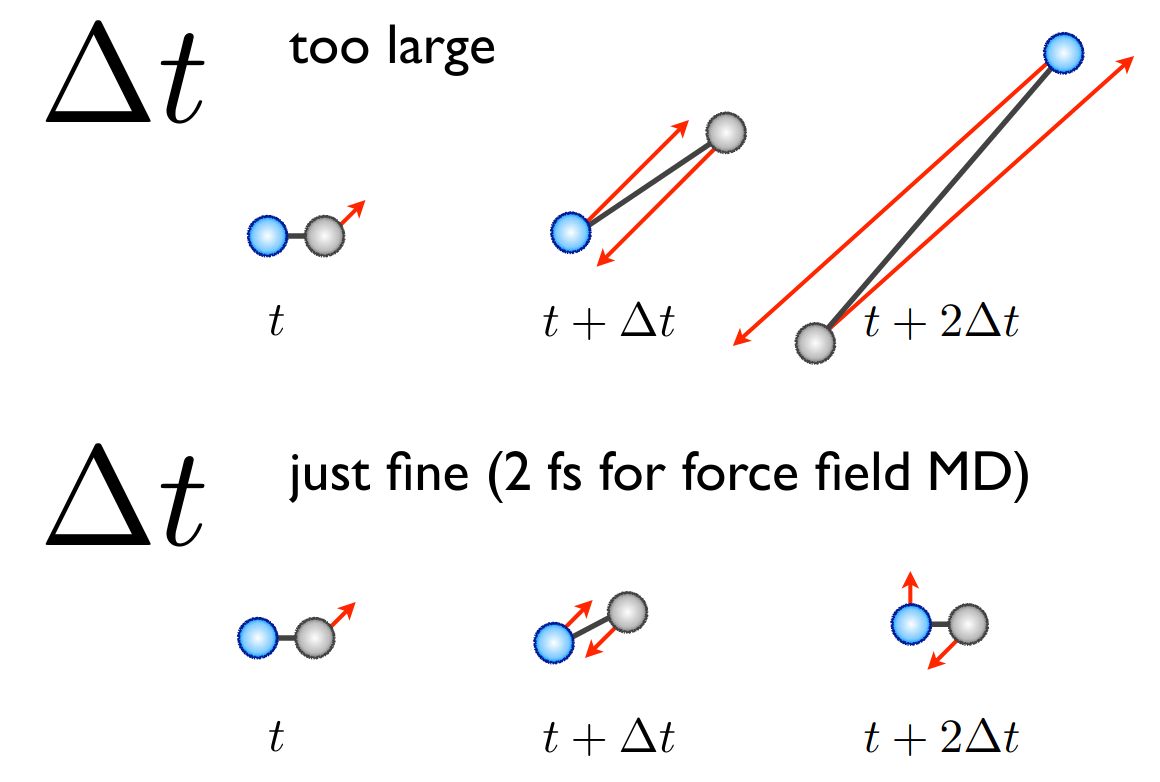}}
\caption{Effect of different time steps on the evolution of two particles (atoms) in a molecular dynamics simulation. The forces acting on the atoms are depicted with the red arrows. Note that if the time step is too large ($>$ 2 fs for full-atomistic MD simulations) the forces increase gradually and the positions of the particles are swapped at each step, preventing a physically accurate description of the dynamics.}
\label{fig:ChMD-TimeStep}
\end{figure}

The time-step $\Delta t$ needs to be chosen carefully. If a too small timestep is used, precious computational effort is wasted. A too large timestep on the other hand will yield integration errors and prevent the proper dynamics of the system, as shown in \figref{ChMD-TimeStep}. Strictly, 10 to 20 integration steps are needed per period of a harmonic vibration to integrate accurately. However, due to Verlet's third-order accuracy, 5 integration steps suffice in practice. Typically, the fastest vibrations in biomolecular simulations, are the hydrogen atoms that vibrate at 10 fs \cite{Feenstra1999}. Therefore, 1 or 2 fs timesteps are typically used for MD simulations. 

To put these time scales in perspective, the 2 fs timestep is at the bottom of the scale in \figref{ChMD:Timescales}. Biologically relevant behaviour starts at the right-hand block; around milliseconds. That means, we need an order of $10^{12}$ integration steps to start getting into biological timescales. This is typically not feasible, however, fortunately there is still a lot of interesting and biologically relevant behaviour that we may observe at achievable timescales of nanoseconds to microseconds.

\begin{panel}
\figlab{A}\includegraphics[width=0.45\linewidth]{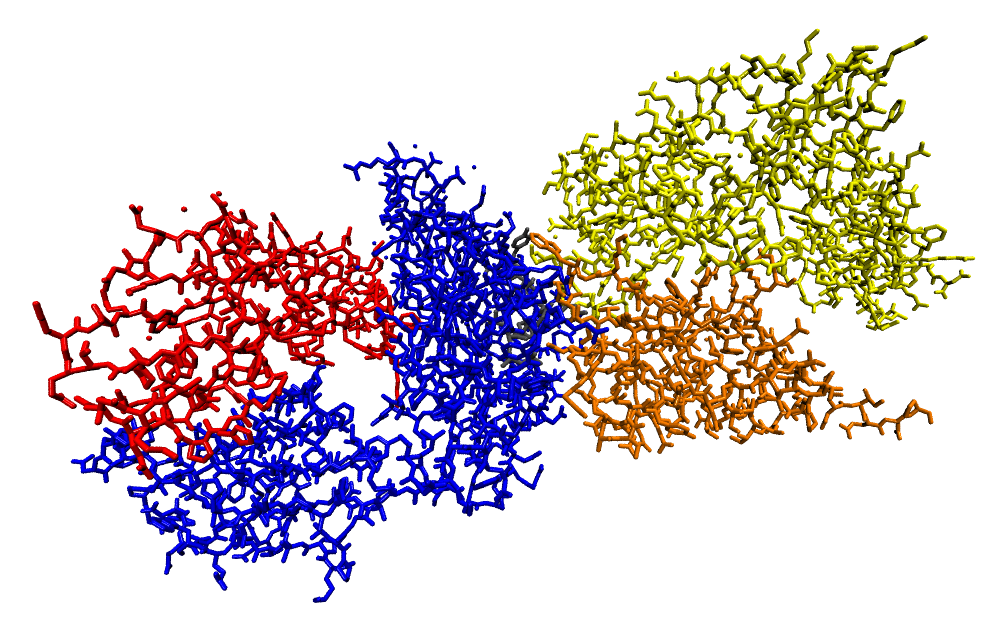}
\figlab{B}\includegraphics[width=0.45\linewidth]{figs/ChMD-protein-water-sticks}
\figlab{C}\includegraphics[width=0.45\linewidth]{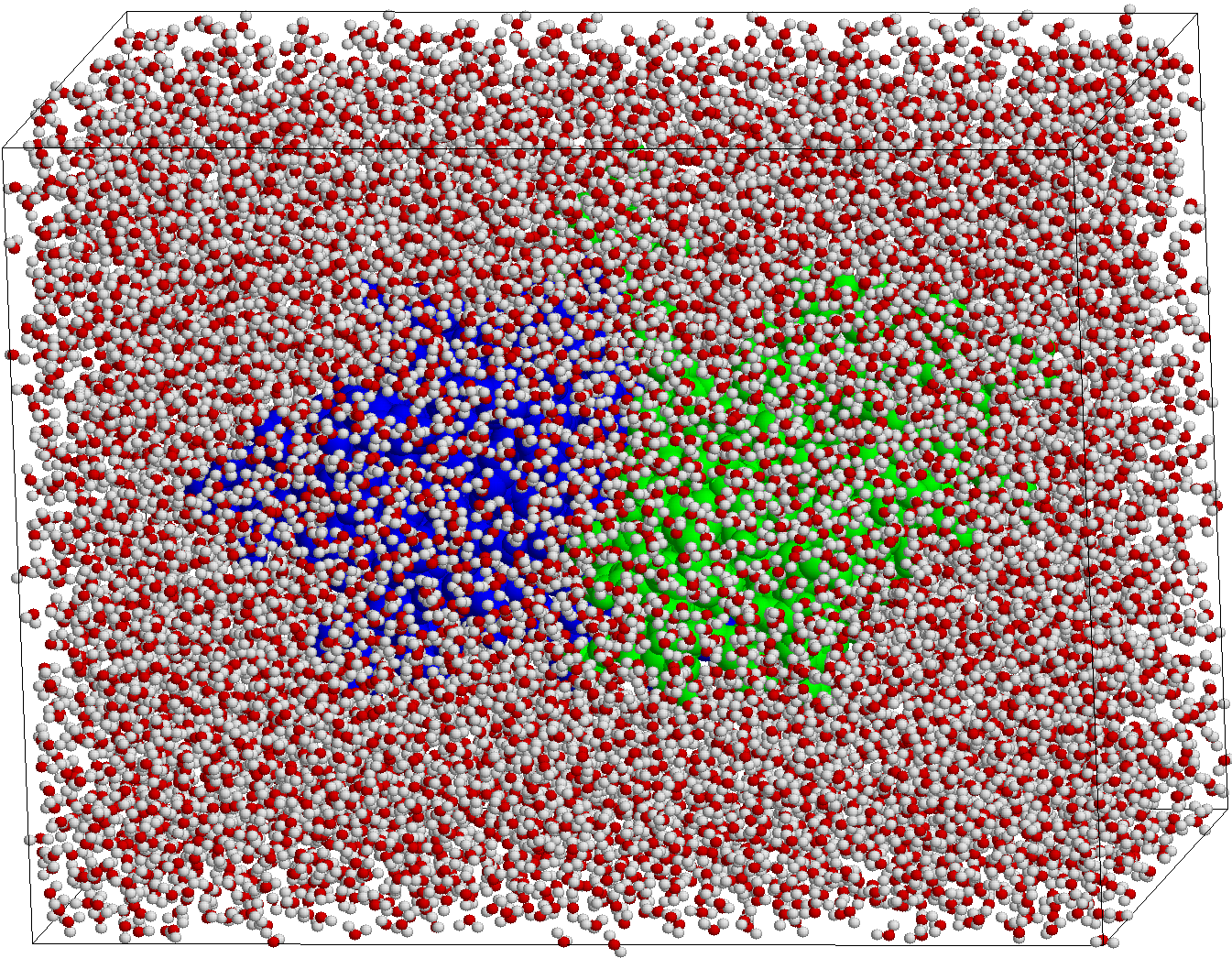}
\figlab{D}\includegraphics[width=0.45\linewidth]{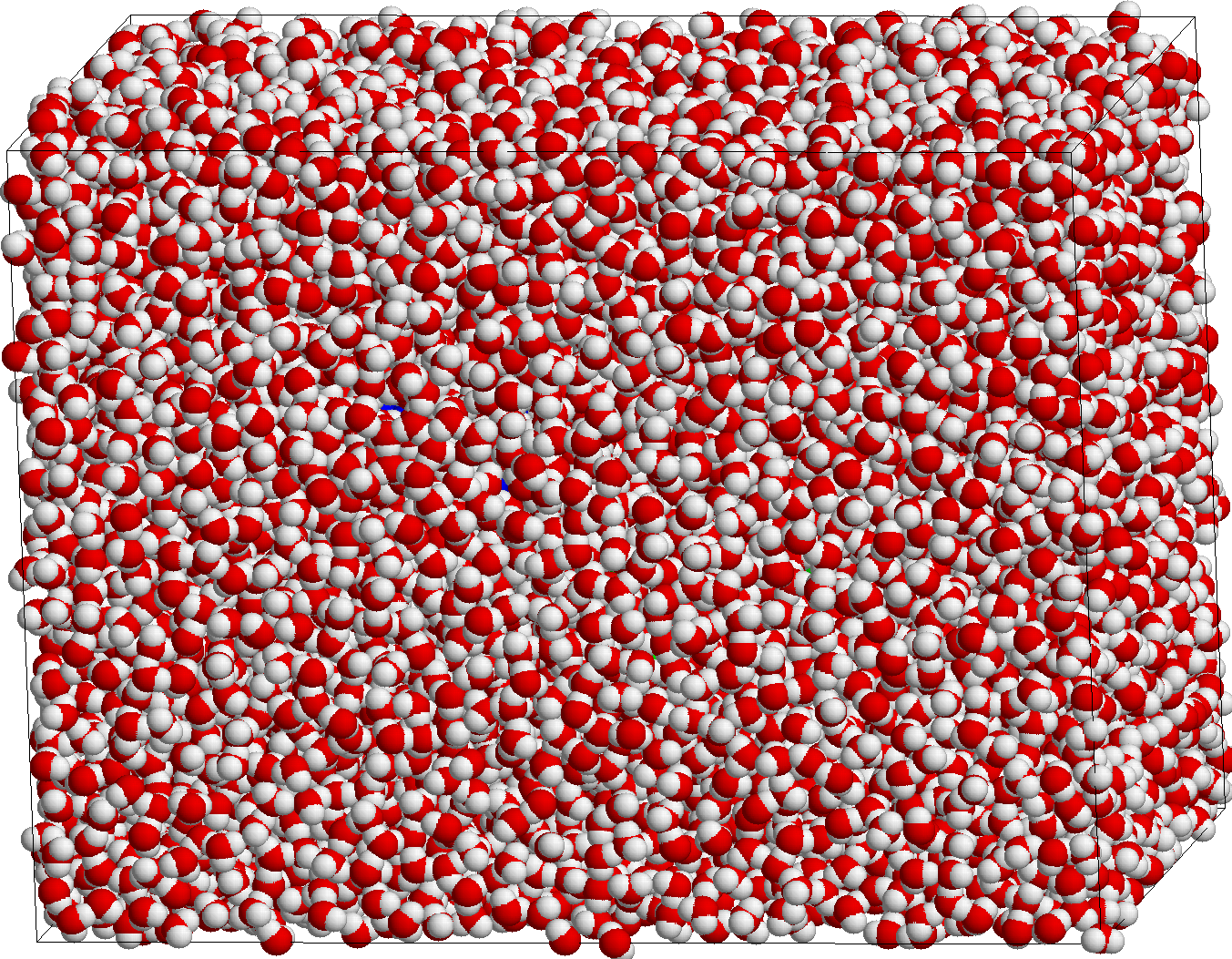}
\caption{Illustration of an exemplary set-ups for an MD simulation. \textbf{A} a protein in vacuum.  \textbf{B-D} solvated in a box with water molecules illustrated as two (O--H) bonds (\textbf{B}),  as small spheres (\textbf{C}), or  using the full Van der Waals radii of the atoms (\textbf{D}).}
\label{fig:ChMD-water}
\end{panel}

\begin{bgreading}[Biological systems: water needed]
When we simulate a protein, the largest part of the investigated system consists of just water. This water is absolutely crucial to the proper behaviour of the protein, as was already introduced in \chref{ChIntroPS} \secref{ChIntroPS:hydrophobic-core}, but not something one is actually interested in by itself. There are two different ways to treat water in a simulation: using either an implicit or an explicit water model. In an explicit water model, the water is modelled as discrete H$_2$O molecules, with a triangular arrangement of the oxygen and two hydrogen atoms. In an implicit model, the water surrounding the protein is described using an average `field' description, which means that water molecules are not treated as actual molecules consisting of particles, but are instead approximated by a dielectric constant which effectively dampens the electrostatic interactions. In the case of explicit water, for two protein molecules of about 3.5 nm diameter, we need a box of 9.3 nm by 6.9 nm by 5.6 nm filled with 10865 water molecules. \figref[A]{ChMD-water} shows the two proteins in vacuum, B shows the proteins with the water molecules drawn as lines. 
In C the water atoms are shown as small spheres where we can still see the protein somewhere in the middle. And in D the full Van der Waals radii of the atoms are drawn which completely hides the protein, thus emphasizing the amount of water needed relative to the size of the protein molecule. This example illustrates why explicit water simulations are much slower and thus computationally more expensive than simulations using an implicit water model. However, depending on the research question and on the experimental counterpart used to validate the simulations, it may be justified to use an implicit model.  
\end{bgreading}

\begin{bgreading}[Common approaches for performance enhancement]
\figref{ChMD:Timescales} highlights the enormous gap between the integration timestep needed, and biologically relevant timescales that we would like to approach in our simulations. So, many tricks have been invented over the years to improve efficiency of the calculations, as well as a wide variety of approximations that enhance performance typically by reducing complexity of the simulated system.

\begin{compactitem}
\item Trivially: The forces derived from pairwise potentials are related as follows $F_{ij} = - F_{ji}$, which means the interaction between an atom pair needs to be computed only once. 
\item The non-bonded potentials (Coulomb: $\sim \sfrac1r$; Lennard-Jones: $\sim \sfrac1{r^6}$ ) tend to zero at large distances, so we can use a distance cut-off beyond which interactions are not calculated. 
      \begin{compactitem}
      \item Atoms only move a little in one step, so we use a pair-list to keep track of atoms within the cut-off distance which we only re-evaluate every so many steps (calculating distances is relatively expensive because there are quadratically many pairs for the number of atoms). 
      \item Evaluating $r$ is relatively expensive due to the square-root; so compare the square of the cut-off radius with the square of the distance. 
      \end{compactitem}
\item Large distances change less, which forms the basis for twin-range and multiple time-step methods. 
\item Many Processor/Compiler/Language specific optimizations:
      \begin{compactitem}
      \item use of Fortran vs. C (and even assembly code) in performance critical parts of the code
      \item optimize cache performance for very large arrays (positions, velocities, forces, parameters)
      \item compiler optimizations
      \item efficient use of multi-core systems
      \item use of GPU nodes for efficient execution of so-called `vector' and grid operations
      \end{compactitem}
\item Maximum Time step is limited by vibrational frequencies (see also \figref{ChMD:Timescales}:
      \begin{compactitem}
      \item The carbon--hydrogen bond angle vibration is $10^{-14}$s = 10 fs. When using 10-20 integration steps per vibrational period, we would need a time step of 0.5 fs, but in practice 1 or 2 fs time steps are used (1.000.000 or 500.000 steps for 1 ns)
      \item Bond distances also have high frequencies of vibration, but for protein simulations these are already always constrained.
      \item If we also remove other fast vibrations (hydrogen atom bond and angle motion and some out of plane motions of aromatic groups) using constraints we can use a timestep of 2 or 4 fs, and even stretch it to 6 or 7 \cite{Feenstra1999,Hopkins2015}.
      \end{compactitem}
\item When using simplified force fields, fewer particles are required, thus reducing the time needed for the force computations. 
      \begin{compactitem}
      \item United atoms: aliphatic groups (a carbon bound to only other carbons and hydrogen atoms) are uncharged and nearly spherical, because the hydrogens are so much smaller than the carbon atoms. These can be treated as single particles with a radius slightly larger than just the carbon atom and a larger mass \cite{Gunsteren1996,Yang_UA_2006}.
      \item Coarse-graining: multiple atoms are combined into one particle with so-called effective interaction potentials, resulting in fewer atoms, and also in (much) larger timesteps \cite{May2014,Singh2019}. 
      \item Implicit water: water may also be described not as a collection of molecules, but instead the water molecules are described as an average `field' \cite{Roux1999,Onufriev2019}.  
      \end{compactitem}
\end{compactitem}
In this list we mention only a few strategies to reduce the simulation time. More comprehensive and technical treatises may be found in \citet{Hess2008} and \citet{Pronk2013}, both focused on the GROMACS MD simulation software, and \citet{Bowers2006}, \citet{Dror2012} and \citet{Grossman2013} from the D.\@ E.\@ Shaw research group who developed their own simulation software but also their own supercomputer based on specialized MD hardware called ANTON.
\end{bgreading}

\subsection{Convergence of state properties}

Viewing the movie resulting from the MD simulation is not an efficient way to analyse simulations, and many details are not registered by the human eye. Moreover, the particular order of events, or even a particular conformation, observed in a simulation, is not of interest in itself. When analysing multiple simulations of the same system, one will observe that the order is not conserved, and many small variations on conformational states exist, due to the chaotic nature of the system, as described in the section on dynamics. 

Thus, in this part, we will introduce how to quantify simulation results.
An important overall feature of molecular simulations is some measure of convergence. Strictly speaking, convergence indicates how well the sampled conformations represent the system at equilibrium.  It can also be seen as a measure of the exploration of the energy landscape (or conformational space) of the system of interest. Moreover, not all parts of the energy landscape are equally relevant to the behavior of the simulated system; (very) high energy parts are (very) unlikely to be visited, in reality and as well as in the simulation, as we saw in the previous \chref{ChThermo}. The parts that do not have (very) high energies are often called the `reachable' conformational space.

So, sampling can be considered as the extent to which this reachable part of conformational space has been visited during the simulation. Finally, complete sampling then corresponds to (complete) convergence. However, you should always keep in mind that the dynamics of proteins are stochastic. When starting a simulation from a given structure you might get stuck in a local minimum, in which case, you will not be able to explore all the relevant regions of your energy landscape. Now, we will show some examples of how to (not) recognize convergence from the results of a simulation.

\subsubsection{Enhanced Sampling techniques}

When launching a molecular dynamics simulation from a given starting point (like a protein crystal structure), there is the danger of getting stuck in a local minimum during the simulations. In that case, you won't be able to sample sufficient parts of the conformational space of your system to study what is going on. To avoid this multiple minima problem, several advanced sampling techniques can be used \cite{Yang_ES_2019}. Here, we want to briefly introduce the concepts of umbrella sampling and replica exchange molecular dynamics (see \panelref{ChMD:Enhanced Sampling}); we will also come back to these in \chref{ChMC}. 

Umbrella sampling requires you to have some prior information of what is going on in your system \cite{Kaestner_US_2011}. 
This is different from REMD which allows you to freely explore the conformational space.
In umbrella sampling, you need to pre-define a specific order parameter (a collective variable) that will properly describe relevant changes in your system. What is a good order parameter depends on your research questions; it may for instance be the distance between a ligand and its protein binding site, or between opposite sides of an opening and closing channel protein, or the end-to-end distance in an unfolding event. Once this is clearly defined, your system will be constrained using harmonic potentials (umbrellas) all along this order parameter to sample the parts of space you are interested in. Similar to REMD, this technique is computationally expensive. However, it does allow you to get a quantitative estimation of the free energy for a given event.      
   
\begin{bgreading}[Replica exchange molecular dynamics (REMD)]
\label{panel:ChMD:Enhanced Sampling}

Replica exchange molecular dynamics (REMD)
allows you to sample large portions of your configurational space by running several identical copies (replicas) of your system (e.g., the protein and water) simultaneously at different temperatures (e.g., between 300 K and 700 K). While the higher temperature replicas are provided with sufficient energy to jump over high energy barriers, the low temperature replicas are able to properly sample the local minima. Over time, neighboring replica are allowed to swap their positions which lets you sample different portions of the surface. The mayor bottleneck of this techniques is is the computational costs due to the fact that all the replica need to be simulated in parallel.  

\end{bgreading}

\subsubsection{Observing sampling -- RMSD}

\begin{figure}
\figlab{A}\includegraphics[width=0.45\linewidth]{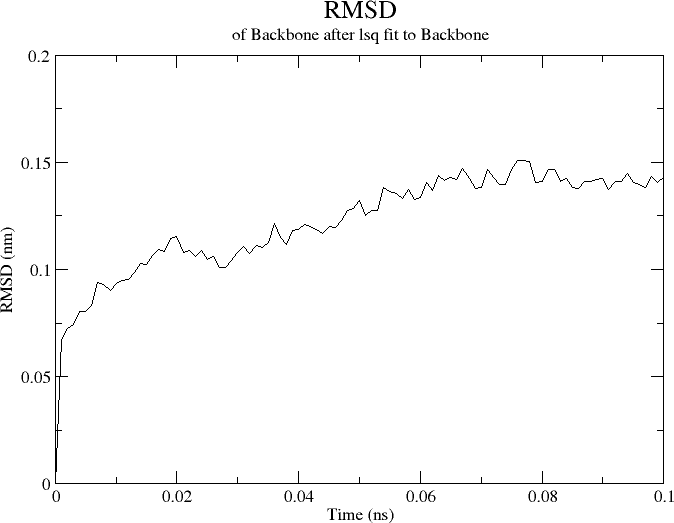}
\figlab{B}\includegraphics[width=0.45\linewidth]{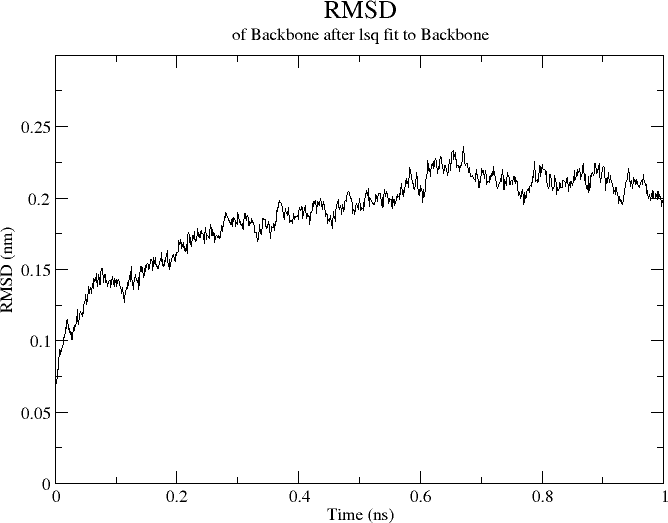}
\figlab{C}\includegraphics[width=0.45\linewidth]{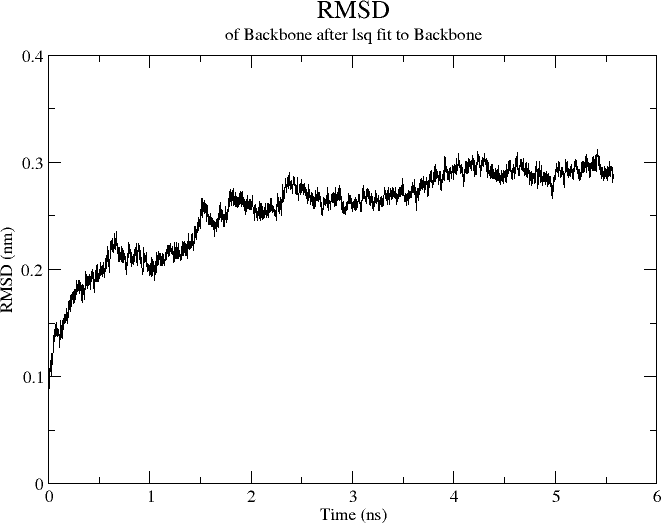}
\caption{Apparent convergence in root-mean-square-deviation (RMSD) may be observed in protein simulations at different timescales. The three plots each derive from the same simulation. At the shortest timescale, up to 0.1 ns or 100 ps \textbf{A}, the RMSD appears to reach a plateau, suggesting convergence. However, extending the simulation to 1 ns \textbf{B}, and to 5.5 ns \textbf{C} shows this convergence is transient.}
\label{fig:ChMD-rmsd-convergence}
\end{figure}

Root mean square deviation (RMSD, introduced in \chref{ChStrucAli}) is one way to chart the progress in sampling the conformational space, by effectively measuring the distance from the starting point. \figref{ChMD-rmsd-convergence} shows the RMSD of the conformation of a protein with respect to the starting (crystal) structure during the simulation. It is clear that the RMSD increases over time, as expected (because the atoms move, the structure changes). Importantly, the three panels show this simulation at three different time scales: 0-100 ps (top left), 0-1 ns (top right) and 0-5.5 ns (bottom right). At each of these timescales, the RMSD appears to reach a plateau roughly half-way through the plot. But the plots are all based on the same simulation, and on the next plot with a longer timescale one sees the increase in RMSD continuing. In other words, each time the plateau we see is `temporary', and the impression it gives of reaching a converged value (in this case for the RMSD) is false. We can be sure this holds for the plateau observed in the bottom right figure as well. If we would run this simulation for much longer, say 100 ns, we would probably still see similar behaviour. These are all symptoms of diffusive motion in a high-dimensional space; as the distance from the starting point increases, the sampling spreads out more and more at a given RMSD distance from the starting point. 
This is similar to diffusion of molecules in a solution where the distance reached from the starting point increases with the square root of the time elapsed -- but the effect is much stronger in high-dimensional conformational space, as we would take a higher power root of the elapsed time here. Therefore, the change in distance gets slower, but the actual changes in conformation still occur as much as in the beginning.

\begin{figure}
\centerline{
\figlab{A}\includegraphics[width=0.6\linewidth]{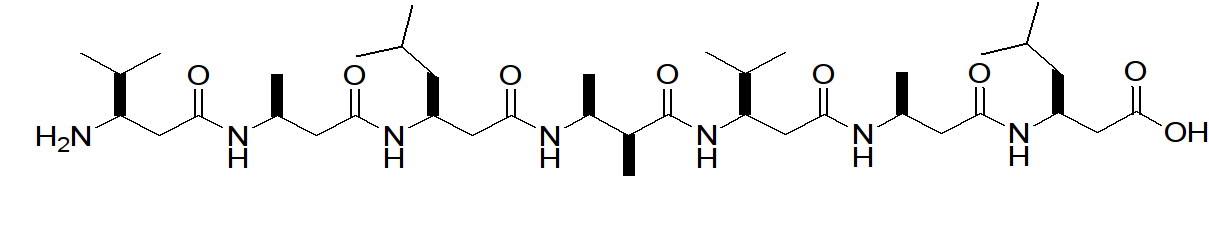}
\figlab{B}\includegraphics[width=0.3\linewidth]{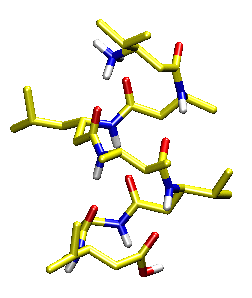}}
\centerline{
\figlab{C}\includegraphics[width=\linewidth]{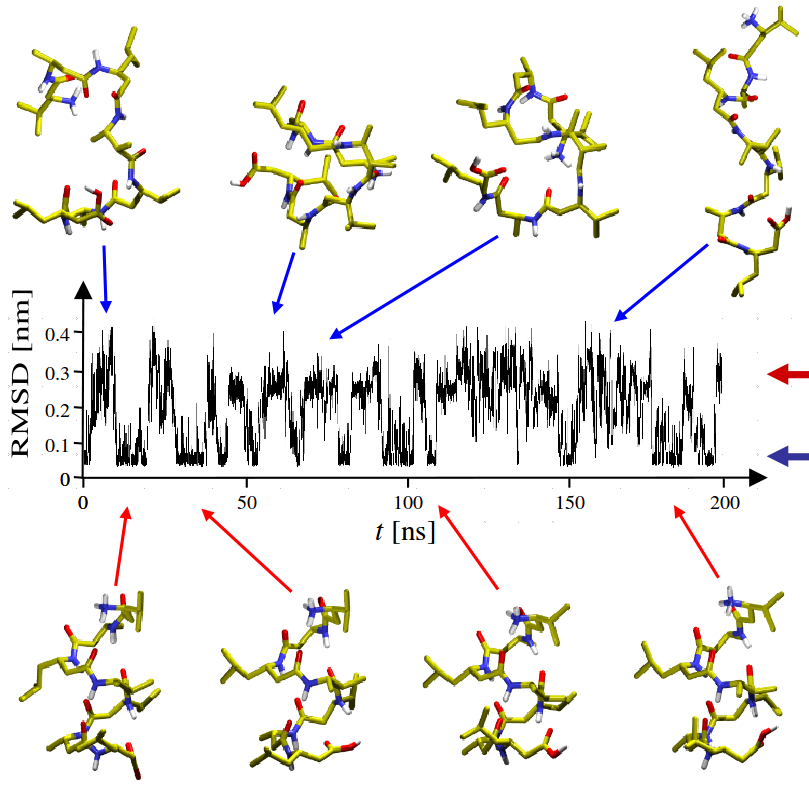}}
\caption{\textbf{A} Small beta-peptide of seven residues (beta-peptides have an additional carbon atom in the backbone compared to normal (alpha) peptides). \textbf{B} The peptide forms a helical structure according to NMR. \textbf{C} Simulations show a very dynamical behaviour, where the RMSD to the helical structure increases, but also decreases again repeatedly. This indicates the peptide unfolding (increasing RMSD) and (re-)folding; importantly this is the very first reversible folding simulation ever \cite{Daura1999,Gunsteren2001}. Reproduced with permission from Daura \& Oostenbrink (pers. comm.).}
\label{fig:ChMD-peptide}
\end{figure}

When sampling of the system reaches timescales at which these conformational transitions start to equilibrate, we should see a change in this behavior. We expect the distances to become smaller again after some time. In fact, \figref{ChMD-peptide} shows that for a small peptide this is achieved at timescales of a few hundred ns.
This seven-residue peptide was studied at the ETH in Z\"urich in the late 1990's. The NMR predicted an $\alpha$-helical conformation for the peptide, shown in \figref[b]{ChMD-peptide}, but the side chains could not be resolved due to lack of data, and some violations of the distances obtained from the experiment. 

Very extensive simulations, for those times, of this peptide were performed in the group of Wilfred van Gunsteren (also ETH Z\"urich) \cite{Daura1999}; note that in those days this was still a major investment of computational power, requiring almost half a year on a large compute cluster. The simulation shows the peptide going through various different conformations, some of them helical (or helix-like), others decidedly different, as can be seen in \figref[c]{ChMD-peptide}. 
The RMSD with respect to the helical (NMR) structure shows that this simulation, for the first time in history, reached equilibrium behaviour. The RMSD not only went up, but also went down again. In other words, during this simulation the system visits different states, but also returns to states visited before. It is important to realize that, in this plot, a low RMSD always means the same state (the native, folded helical state, as can be seen in the snapshots below the plot), but a high RMSD can mean many different things. The snapshots above the plot show that these states are all clearly not helical, but they are also not similar to each other. Clustering of these unfolded structure showed that there are (only) about 1000 distinct conformational states in this unfolded ensemble (for details, please refer to \citealt{Gunsteren2001}).

Ten years later in 2010, Shaw et al. showed that now we can obtain reversible folding also for a  small (70 residue) protein. The timescale needed is much larger than that for the peptide; here we see the simulation goes up to 200 $\mu$s (note, that this is a similarly major investment of computational power as the peptide was a decade earlier, of the order of many months in parallel on a large number of highly specialized CPU's).

\begin{bgreading}[Evaluating your MD simulations -- Order Parameters]
\label{panel:ChMD:Analysis}
To be able to track the progress of a molecular dynamics (MD) simulation, several powerful analysis tools have been developed over the years. Although it may be fun and also seemingly intuitive to visualize your results as a movie by loading the MD trajectories into programs such as VMD or Chimera, it is not sufficient to understand what is going on and you need a quantitative way of analysing your results.   
Luckily, MD packages provide a large number of ready-implemented parameters that can be used to track changes between states of interest in your simulations. First, you should check the general properties of your system, such as the average pressure and the temperature, to make sure everything went well. In a next step, tracking the changes of the following \emph{order parameters} may help you pinpoint potential events during your simulations:   
\begin{compactitem} 
\item \textbf{Root Mean Square Deviation}: Measure of dissimilarity (distance) between two molecular conformations used to compare the conformation of each frame with respect to, e.g., the starting point of the simulation
\item \textbf{Root Mean Square Fluctuation}: Measure of structure variation, it calculates the standard deviation of the deviation of an atomic position over time.
\item \textbf{Radius of gyration}: Measure for the compactness of a structure, powerful to identify unfolding event in your protein structures. 
\item \textbf{Solvent Accessible Surface Area}: Measure to determine the surface of the biomolecules that are accessible to the surrounding solvent (usually water). This is very helpful in protein folding and stability studies.
\item \textbf{Number of hydrogen bonds}: Compute the numbers of hydrogen bond contacts in your system. Particularly interesting when looking at complexes (ligand-protein, protein-protein) and (un)folding events. 
\item \textbf{Essential Dynamics} (ED) analysis \cite{amadei93,aalten97}:
This technique is a principal components analysis (PCA) of the atomic coordinates. It takes the covariance matrix of the coordinates of the atoms. This covariance matrix is diagonalized, yielding eigenvalues and eigenvectors; the eigenvectors describe collective motions of the atoms analyzed. The eigenvalues (or loadings) represent the magnitude of each of these motions. We can now simply focus analysis on the eigenvector(s) with the largest eigenvalues, or several of the largest -- these are also referred to as `Essential Modes'. Finally, if we choose only the two largest, we can project motions onto a two-dimensional plot for visualization.
\end{compactitem}
\noindent
For all these measures, the parameter of interest does not have to be calculated for the whole system -- often for example only the C$\alpha$-atoms or the active site or binding region are analyzed.
 
Depending on the problem at hand, looking at additional order parameters may be interesting. For example to analyse the simulation of a ligand-protein complex, the contact frequency of the ligand with specific amino acid inside the binding site or the distance between the centers of mass. Finally, despite the large number of available tools that come with most simulation packages, in some cases, your simulations may require you to do some self-coding to be able to follow important changes during your simulations. A Python library to analyse your simulations that may come handy can be found here: \url{https://www.mdanalysis.org/} \cite{Michaud-Agrawal2011,Gowers2016}.

\end{bgreading}

\subsection{Temperature dependence}
In the previous \chref{ChThermo}, \figref[a,b]{ChThermo:reversible-peptide-temperature}, we already saw how the folding-unfolding equilibrium that we are sampling in these simulations for this peptide, depends on the temperature of the simulation. The middle plot (at 340 K) is the one also shown in \figref[c]{ChMD-peptide}, where on average the peptide is folded 50\% of the time (and unfolded the other 50\%). At lower temperatures in \figref[a]{ChMD-peptide-TP-dep}, we see the peptide tends to spend a larger fraction of time in the folded state. At higher temperatures we see the opposite; a larger fraction of time spent in the unfolded state(s). Note that the temperature changes are fairly small, only 10-20K up, and 20-40K down, and it is reassuring that simulations are able to capture these relatively subtle effects. \figref[b]{ChMD-peptide-TP-dep} shows that also high pressure has the effect of unfolding. Both effects of high temperature and high pressure unfolding of proteins are also experimentally well known facts, for by far the majority of proteins.

\begin{figure}
\centerline{
\figlab{A}\includegraphics[width=0.8\linewidth]{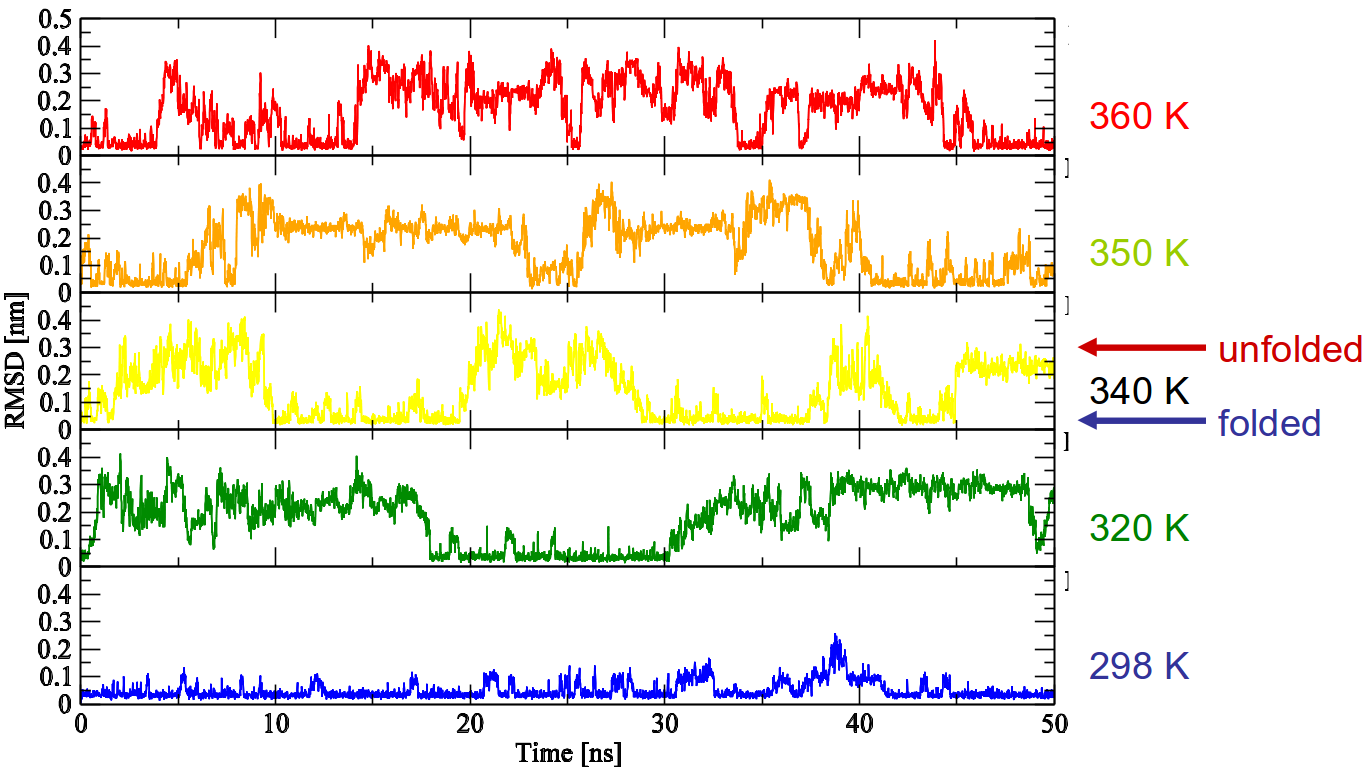}}
\centerline{
\figlab{B}\includegraphics[width=0.8\linewidth]{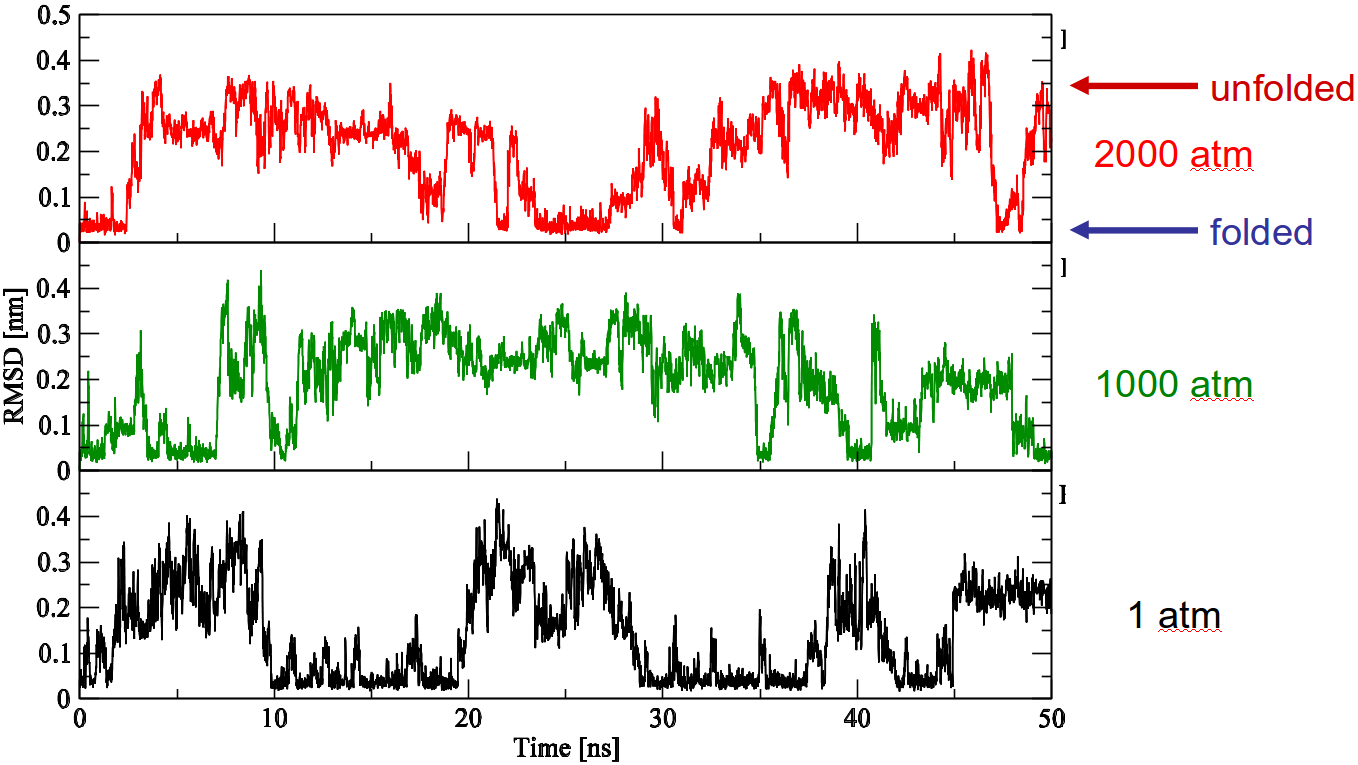}}
\caption{\textbf{A} The folding equilibrium of the beta-peptide depends on temperature: at lower temperatures a larger fraction of time is spent in the folded (low RMSD) state, at higher temperatures a smaller fraction is folded. \textbf{B} Also at higher pressure, the fraction folded decreases \cite{Daura1999,Gunsteren2001}. Reproduced with permission from Daura \& Oostenbrink (pers. comm.).}
\label{fig:ChMD-peptide-TP-dep}
\end{figure}

\subsubsection{Simulated annealing}

One way to overcome free energy barriers and escape from free energy minima in a molecular system is to increase the temperature. At room temperature a protein system can be trapped in a free energy minimum from which it can escape only after a very long time, as we already discussed previously. For MD simulations, a microsecond is already considered a long time. When performing a simulation at higher temperature, 400 K for example, the system can explore many more different configurations than at room temperature. 
However, such configurations do not necessarily represent configurations of the system at room temperature, as then they might be very unlikely. To overcome this, one can take a number of snapshots from the high temperature simulation, and use these as a starting point for MD simulations at room temperature. The snapshots will then equilibrate to configurations that are likely to occur at room temperature. This approach is known as simulated annealing and is used in for example resolving NMR structures by letting high temperature conformations anneal to the constraints obtained from various NMR spectra. 

One problem with the simulated annealing approach is that the resulting conformations do not represent the equilibrium distribution. The selection of the snapshots chosen to equilibrate to room temperature does not follow the Boltzmann distribution. In principle, the MD simulation at room temperature should sample the equilibrium distribution, but may require a long time to achieve that. To overcome this issue, but still make use of the enhanced sampling provided by performing MD at high temperature, the procedure to select snapshots for equilibration at lower temperature should follow the Boltzmann distribution. The solution to this is parallel tempering or replica exchange, which we already touched on in \panelref{ChMD:Enhanced Sampling}.

\subsection{Homology model optimization}
Now we will illustrate simulation, sampling and convergence in practice. Here we have used MD simulations to optimize details of a homology model of an enzyme \cite{Feenstra2006}. \figref[a]{ChMD:SMO} is an overview picture of the enzyme Styrene mono-oxygenase (SMO) homology model protein structure. If you look closely, you will observe two additional molecules in the center: the styrene (STY) ligand, and the flavin-adenin-dinucleotide (FAD) co-factor. 

The homology model for this structure was built based on two templates of about 23\% sequence identity, which is far enough to be a difficult target for homology modelling. Docking of the ligand into the model yielded binding conformations that did not correspond to known products of this enzyme. We therefore used MD simulations to optimize the model, with a main aim to improve the shape of the active site pocket, but to do that we would also need to relax strain due to bad contacts in the initial model. The procedure was therefore aimed to relax this strain, without leading to distortion of the binding site region. We first allowed the water to relax around the protein during a 1-ps MD where the positions of the atoms in the protein were restrained. Subsequently, we released the constraints on parts of the protein, so these were also allowed to relax. First, only the side chains of the residues outside the binding pocket were released, and simulated for 1 ps. Next, also the backbone of these residues were released and simulated for 10 ps. Then, the side chains of the binding residues were released, and simulated for another 10 ps. During this whole procedure the positions of the FAD co-factor and the styrene substrate were also restrained, this allowed the protein structure to relax around the bound co-factor and substrate. Finally, only the backbone atoms of the binding residues were restrained for 100 ps, while the rest of the protein as well as FAD and styrene were free.

To be able to track progress during the optimization simulations for this homology model, we used Essential Dynamics (ED) analysis \cite{amadei93,aalten97}, see also \panelref{ChMD:Analysis}. We can now simply focus analysis on the two largest eigenvalues (the essential modes) and project these motions onto a two-dimensional plot for visualization.

\begin{figure}
\centerline{
\figlab{A}\includegraphics[width=\linewidth]{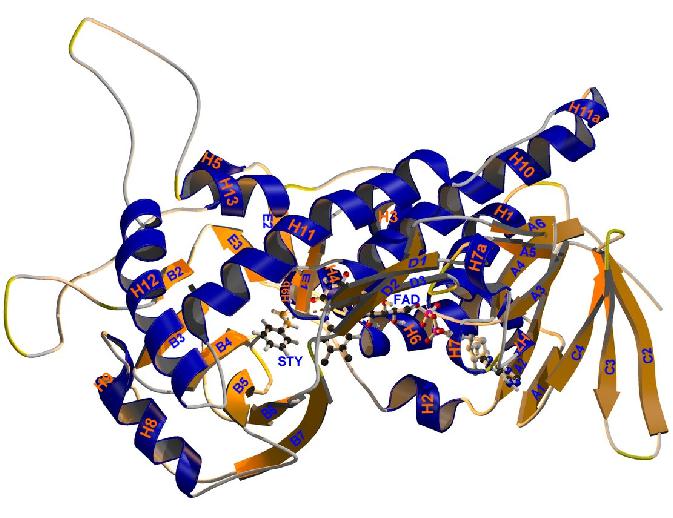}}
\figlab{B}\includegraphics[width=0.45\linewidth]{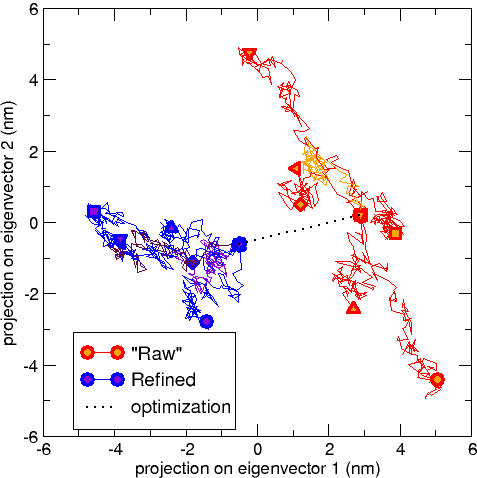}
\figlab{C}\includegraphics[width=0.45\linewidth]{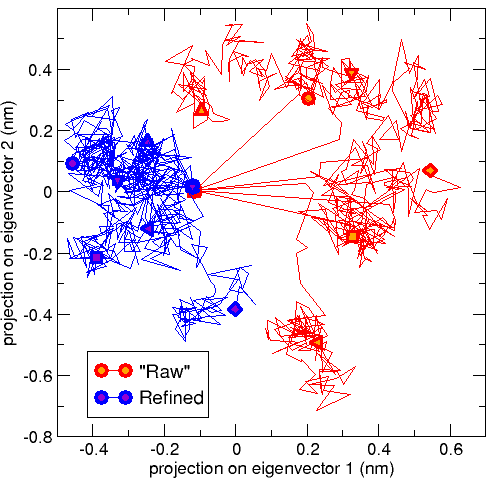}
\caption{\textbf{A} Structure of the Homology Model Styrene Mono-Oxygenase (SMO) Enzyme. \textbf{B} Essential Dynamics (ED) analysis of the C$\alpha$ atoms, showing backbone rearrangements during simulations starting from the `raw' structure (blue), which are distinct from those started from the `refined' homology model (red). The dotted line indicates the optimization path. \textbf{C} The same, but now the ED analysis was performed on the active site region only. Here, structural effects (difference between starting points) are small - the optimization path can not even be seen here. Nevertheless, overall behaviour of the `refined' simulations is still distinctly different from that of the `raw' simulations. The long straight lines `shooting' out from the `raw' starting point indicate high levels of strain in the `raw' structure (red), which is relaxed in the refined structure (blue). Figure modified from \citet{Feenstra2006}.}
\label{fig:ChMD:SMO}
\end{figure}

To assess stability of the protein, and to see if the optimization procedure was successful in improving stability, we performed two sets of simulations, one started from the non-optimized `raw' homology model, while the second set of simulations was started from the optimized model. We visualized the behaviour in these two sets of simulations using ED analysis on both. \figref[b]{ChMD:SMO} shows the ED plot representing the motion during these simulations, projected onto the two largest eigenvectors; i.e.\@ the two largest collective motions going on in these simulations. It is clear, that both sets of simulations (`raw' vs. `optimized') have different behaviour. The `raw' simulations tend to start out by moving rapidly away from their starting point, and in different directions; this indicates strain in the initial conformation. In contrast the `optimized' simulations move more gradually and more similarly. Finally, it is clear that both sets of simulations have a different starting point: this of course corresponds to the changes in the structure due to the optimization procedure.

The motion of the protein structure corresponding to each of these two eigenvectors can be visualized in a movie. In such a movie, you will be able to see that all atoms contribute to the overall motion, although not all atoms equally. You will also see that the two (eigenvector) motions are distinct; different atoms contribute more to each of them, and the motions are (often) in different directions. Finally, you will also see that different parts, sometimes quite far away in the protein structure, move in and out together in a sort of concerted way.

\section{Outlook and summary}
Molecular dynamics simulations have shown their usefulness in giving a detailed view on how proteins work; in the words of Shaw c.s., a ``Computational Microscope for Molecular Biology''  \cite{Dror2012}. With ever increasing computational power, microsecond simulations are within reach and millisecond simulations have already been produced for a few selected systems. Current challenges lie in making available methods accessible, and in handling and analysing the huge amounts of data produced from the simulations. Before starting to run an MD simulation, there are two important things that you need to think about, namely which simulation technique is adequate to solve your scientific question and do you have experimental data to validate your models. As numerous techniques have been developed over the years and are now available to the scientific community, you should make sure to understand the basic concepts, usefulness and limitation of the technique you want to use. Later, it is extremely important to support your results and validate your models to guarantee their physical relevance. Often, you will be able to work in close collaboration with experimental groups or even perform experiments yourself. Alternatively, experimental data can be used from published results in the literature. In any case, make sure to be aware of what is going on in the experiment, the information it is able to provide, its benefits, bottlenecks, errors sources, and limitations.

\section{Key concepts}
Concepts that should be known after reading this chapter:
\begin{compactitem}
\item The importance of molecular motions:
\begin{compactitem}
\item Proteins are dynamic and understanding their motions is crucial to be able to understand their biological function
\item Always remember and be aware of the relevant length and time scales to observe a biological event
\end{compactitem}
\item MD Simulations are based on:
\begin{compactitem}
\item Newton's classical mechanics
\item force fields that contain the description (parametrization) of the bonded and non-bonded interactions
\item integrating the equations of motion
\item describing thermodynamics ensembles of the system
\item Trajectory on Energy Surface
\end{compactitem}
\item In protein Dynamics:
\begin{compactitem}
\item make sure that you have an adequate experimental counterpart to guide and validate your simulations
\item the level of detail of your forcefield, e.g.\@ atomistic or coarse-grained, should match the problem you are interested in
\item attainable timescales and level of convergance should match the problem you are interested in
\end{compactitem}
\end{compactitem}
\section{Further reading}
\begin{compactitem}
\item Understanding Molecular Simulation -- \citet{FrenkelSmit} 
      \begin{compactitem}
        \item [$\rightarrow$] General, in-depth introduction to simulation
      \end{compactitem}
\item Simulating the Physical World -- \citet{Berendsen2007}
      \begin{compactitem}
        \item [$\rightarrow$] Technical treatise on physical properties, simulation algorithms and statistical thermodynamics
      \end{compactitem}
\item Molecular Modelling: Principles and Applications --
\citet{Leach}
      \begin{compactitem}
        \item [$\rightarrow$] Overview of Molecular Modelling and Computational simulations.
      \end{compactitem}
\item The Art of Molecular Dynamics Simulation -- \citet{Rapaport2004}
\end{compactitem}

\section*{Author contributions}
{\renewcommand{\arraystretch}{1}
\begin{tabular}{@{}ll}
\ACtxt: &   HM, JvG, QH, AM, JV, KAF \\
\ACfig: &   HM, ASR, JV, KAF \\
\ACref: &   HM, ASR, JV, KAF \\
\ACproof:&  HM, JvG, ASR, JV, SA \\
\ACfb:  &   QH, AM \\
\ACeds: &   HM, SA, KAF
\end{tabular}}

\mychapbib
\clearpage

\cleardoublepage

\end{document}